\documentclass[sn-mathphys]{sn-jnl}
\jyear{2021}
\usepackage{caption}
\usepackage{subcaption}
\usepackage{enumerate}
\usepackage{amsmath}
\usepackage{amsthm}
\usepackage{booktabs}
\usepackage{diagbox}
\usepackage{multicol}
\usepackage{multirow}
\usepackage{bm}
\theoremstyle{thmstyleone}
\newtheorem{theorem}{Theorem}

\theoremstyle{thmstyletwo}
\newtheorem{remark}{Remark}
\theoremstyle{thmstylethree}
\newtheorem{definition}{Definition}

\newcommand{\rn}{\mathbb{R}^L}

\newcommand{\inte}{\mathbb{Z}_L}
\newcommand{\complex}{\mathbb{C}^L}

\begin{document}

\title[Star DGT: a Robust Gabor Transform for Speech Denoising]{Star DGT: a Robust Gabor Transform for Speech Denoising}

\author*[1]{\fnm{Vasiliki} \sur{Kouni}}\email{vicky-kouni@di.uoa.gr}
\author[2]{\fnm{Holger} \sur{Rauhut}}\email{rauhut@mathc.rwth-aachen.de}
\author[1]{\fnm{Theoharis} \sur{Theoharis}}\email{theotheo@di.uoa.gr}

\affil*[1]{\orgdiv{Department of Informatics and Telecommunications}, \orgname{National and Kapodistrian University of Athens}, \orgaddress{\city{Athens}, \country{Greece}}}
\affil[2]{\orgdiv{Chair for Mathematics of Information Processing}, \orgname{RWTH Aachen University}, \orgaddress{\city{Aachen}, \country{Germany}}}

\abstract{In this paper, we address the speech denoising problem, where Gaussian, pink and blue additive noises are to be removed from a given speech signal. Our approach is based on a redundant, analysis-sparse representation of the original speech signal. We pick an eigenvector of the Zauner unitary matrix and --under certain assumptions on the ambient dimension-- we use it as window vector to generate a spark deficient Gabor frame. The analysis operator associated with such a frame, is a (highly) redundant Gabor transform, which we use as a sparsifying transform in denoising procedure. We conduct computational experiments on real-world speech data, using as baseline three Gabor transforms generated by state-of-the-art window vectors in time-frequency analysis and compare their performance to the proposed Gabor transform. The results show that our proposed redundant Gabor transform outperforms all others, consistently for all signals and all examined types of noise.}


\maketitle

\section{Introduction}\label{sec1}
Noise is one of the main factors that affect the accuracy of the results in audio processing. This explains why audio denoising is one of the most extensively studied inverse problems in signal processing. The task consists in recovering an audio signal $x\in\rn$ from corrupted linear observations
\begin{equation}
    y=x+e\in\rn.
\end{equation}
Noise removal from audio signals is an important first step in applications such as sound classification \cite{soundclass}, sound event localization \cite{sel}, speech recognition \cite{whisper}, dereverberation \citep{reverb}, speech enhancement \cite{enhance} and source separation \cite{soursep}.
\subsection{Related Work}
In order to address the denoising problem, numerous approaches have emerged, including statistical models \cite{bayes,probability}, empirical mode decomposition \citep{emd}, spectral subtraction \cite{specsub,colorsubtract}, thresholding methods \cite{mallat,durfler}, neural networks \cite{wavenet,relu}, sparse and redundant representations \cite{plumbley,lowrank}, or a combination of the aforementioned approaches \citep{threshemd,aegan}. Sparse and redundant representations have shown very promising results \cite{fletcher,thresh}, especially when turning to analysis sparsity (also known as co-sparsity) \cite{audascity,genzel,figu}, which provides flexibility in modelling sparse signals, since it leverages the redundancy of the involved analysis operators. Albeit the authors of \cite{genzel} are mainly focused on analysis Compressed Sensing \cite{rauhut}, they state that using the analysis-prior formulation --with a redundant analysis operator-- in denoising (from now on we shall call this framework \textit{analysis denoising}) is fundamentally different from classical denoising via soft thresholding. Similarly, \cite{figu} explores the superior reconstruction produced by analysis denoising of 1D signals over its synthesis counterpart \cite{elad}.

\subsection{Motivation}
Our work is inspired by the articles \cite{mallat,genzel,figu,durfler,greedy,pfander}. These publications propose either analysis operators associated with redundant frames (i.e. matrices whose atoms/rows form a frame of the ambient space) with atoms in general linear position, or a finite difference operator (associated to the popular method of total variation \cite{tv}), in which many linear dependencies appear for large dimensions. Moreover, in \cite{mallat,durfler,greedy}, Gabor transforms are combined with thresholding methods for audio denoising. In a similar spirit, we also deploy frames, but we differentiate our approach by using \textit{spark deficient Gabor frames} under the \emph{analysis denoising formulation}. The elements of spark deficient frames are not in general linear position, while analysis denoising differentiates itself --as already stated-- from classical thresholding denoising. Our intuition behind employing spark deficient frames is based on remarks of \cite{cosparse}. In the latter, the authors state that according to the union-of-subspaces model \cite{union}, it is desired to have analysis operators that exhibit high linear dependencies among their rows; this is a condition satisfied by spark deficient frames. Particularly, we choose spark deficient Gabor frames instead of other classes of frames that may be spark deficient (e.g. equiangular tight frames \citep{equiangular}) due to the fact that time-frequency representations resemble the way that the human auditory system works \citep{auditory,intensity};
thus, they are better suited to the application of speech denoising. To that end, we take advantage of an analysis operator \cite{star}, namely star digital Gabor transform (star-DGT), associated with a spark deficient Gabor frame (SDGF). The latter can be generated\footnote{such a frame can also be generated by the eigenvectors of certain unitaries belonging to the Clifford group} by time-frequency shifts of any eigenvector of the Zauner unitary matrix \cite{zauner}, under certain assumptions on the signal's dimension. To the best of our knowledge, the efficiency of star-DGT when applied to denoising has not yet been demonstrated. Therefore, it is intriguing to compare the robustness of our proposed Gabor analysis operator to three other Gabor transforms, emerging from state-of-the-art window vectors, by applying all four of them to analysis denoising. Finally, we illustrate the practical importance of our method for real-world speech signals.
\subsection{Key Contributions}
Our novelty is twofold: 
(a) we generate a SDGF based on a window vector, associate this SDGF to a highly redundant Gabor analysis operator and use the latter as a sparsifying transform in analysis denoising (b) we compare numerically our proposed method with three other Gabor analysis operators, based on common windows of time-frequency analysis, on real-world speech data, arguing also about the selection of the lattice parameters. Our experiments show that our method outperforms all others, consistently for all speech signals and all types of noise -- Gaussian, pink and blue.
\subsection{Paper organization}
The rest of the paper is outlined as follows. In Section 2, we give notation and briefly present the setup of analysis denoising. Section 3 introduces Gabor frames and extends to spark deficient ones, building the desirable SDGF and its associated analysis operator. In Section 4, we describe the experimental settings, while in Section 5, we present two sets of experiments, with corresponding results and evaluation. Lastly, in Section 6 we make some concluding remarks and give potential future directions.

\section{Gabor denoising setup}
\subsection{Notation}
\begin{itemize}
    \item For a set of indices $N=\{0,1,\dots,N-1\}$, we write $[N]$.
    \item (Bra-kets) The set of (column) vectors $\lvert0\rangle,\lvert1\rangle,\dots,\lvert L-1\rangle$ is the standard basis of $\mathbb{C}^L$.
    \item We write $\inte$ for the ring of residues $\mathrm{mod}L$, that is $\inte=\{0\mathrm{mod}L,1\mathrm{mod}L,\dots,(L-1)\mathrm{mod}L\}$.
    \item We write $a\equiv b(\mathrm{mod}L)$ for the congruence modulo, where $a,\,b\in\mathbb{Z}$.
    \item The support of a signal $x\in\rn$ is denoted by $\mathrm{supp}(x)=\{i\in[L]:x_i\neq0\}$. For its cardinality, we write $\lvert\mathrm{supp}(x)\rvert$ and if $\lvert\mathrm{supp}(x)\rvert\leq s<<L$, we call $x$ $s$-sparse.
\end{itemize}

\subsection{Analysis denoising formulation}
As we mentioned in Section 1, the main idea of speech denoising is to reconstruct a speech signal $x\in\rn$ from
\begin{equation}
    y=x+e\in\mathbb{R}^L,
\end{equation}
where $e\in\mathbb{R}^L$, $\|e\|_2\leq\eta$, corresponds to noise. To do so, we first assume there exists a redundant sparsifying transform $\Phi\in\mathbb{C}^{P\times L}$ ($P>L$) called the analysis operator, such that $\Phi x$ is (approximately) sparse. This is an \textit{analysis sparsity model} for $x$.

On the other hand, the type of noise $e$ that is added on $x$ depends on the speech recording method. Noise may originate from the microphones or the environment, in the form e.g. of white noise, pink noise or babble noise. In this paper, we examine three types of additive noises: zero-mean Gaussian with standard deviation $\sigma$, pink and blue.

Using analysis sparsity in denoising, we wish to recover $x$ from $y$. A common approach is the \textit{analysis basis pursuit denoising} problem
\begin{equation}
    \label{denl1}       \min_{x\in\rn}\|\Phi x\|_1\quad\text{subject to}\quad \|x-y\|_2\leq \eta,
\end{equation}
or a regularized version\footnote{in terms of optimization, it is preferred to solve \eqref{regl1} instead of \eqref{denl1}} \cite{tfocs} of it:
\begin{equation}
\label{regl1}
    \min_{x\in\rn}\|\Phi x\|_1+\frac{\mu}{2}\|x-x_0\|_2^2\quad\text{subject to}\quad \|x-y\|_2\leq \eta,
\end{equation} where $x_0$ denotes an initial guess on $x$, $\mu>0$ is a smoothing parameter and $\eta>0$ an estimate on the noise level.

We will devote the next Section to the construction of a suitable analysis operator $\Phi$.

\section{Gabor frames}
\subsection{Gabor systems}
A \textit{discrete Gabor system} $(g,a,b)$ \cite{dgs} is defined as a collection of time-frequency shifts of the so-called window vector $g\in\mathbb{C}^L$, expressed as
\begin{align}
    \label{gaborsystem}
    g_{n,m}(l)&=e^{2\pi imbl/L}g(l-na),\quad l\in[L],
\end{align}
where $a,\,b$ denote time and frequency parameters (also known as lattice parameters) respectively, $n\in[N]$ chosen such that $N=L/a\in\mathbb{N}$ and $m\in[M]$ chosen such that $M=L/b\in\mathbb{N}$ denote time and frequency shift indices respectively. If \eqref{gaborsystem} spans $\mathbb{C}^L$, it is called \textit{Gabor frame} and an equivalent definition of a frame \cite{mal} is given below.
\begin{definition}
Let $L\in\mathbb{N}$ and $(\phi_p)_{p\in P}$ a finite subset of $\mathbb{C}^L$. If the inequalities
\begin{equation}
    c_1\|x\|_2^2\leq\sum_{p\in P}\lvert\langle x,\phi_p\rangle\rvert^2\leq c_2\|x\|_2^2
\end{equation}
hold true for all $x\in\complex$, for some $0<c_1\leq c_2$ (frame bounds), then $(\phi_p)_{p\in P}$ is called a \textit{frame} for $\complex$.
\end{definition}

\begin{remark}
The number of elements in $(g,a,b)$ according to \eqref{gaborsystem} is $P=MN=L^2/ab$ and if $(g,a,b)$ is a frame, we have $ab<L$ (the so-called \emph{oversampling} case). A crucial ingredient in order to have good time-frequency resolution of a signal with respect to a Gabor frame, is the appropriate choice of the time-frequency parameters $a$ and $b$. Apparently, this challenge can only be treated by numerically experimenting with different choices of $a,\,b$ with respect to $L$. In the following subsection, we associate two operators to a Gabor frame.
\end{remark}

\subsection{The analysis and synthesis operators associated with a Gabor frame}
\begin{definition}
Let $\Phi_g:\complex\mapsto\mathbb{C}^{M\times N}$ denote the \textit{Gabor analysis operator} --also known as DGT\footnote{so we will interchangeably use both terms from now on}-- whose action on a signal $x\in\complex$ is defined as
\begin{equation}\label{gabcoeff}
    c_{m,n}=\sum_{l=0}^{L-1}x_l\overline{g(l-na)}e^{-2\pi imbl/L},
\end{equation} for $m\in[M],\,n\in[N]$.
\end{definition}
\begin{definition}
The adjoint of the analysis operator defined in \eqref{gabcoeff}, is the \textit{Gabor synthesis operator} $\Phi_g^T:\mathbb{C}^{M\times N}\mapsto\complex$, whose action on the coefficients $c_{m,n}$ gives
\begin{equation}
    \Phi_g^Tc_{m,n}=\sum_{n=0}^{N-1}\sum_{m=0}^{M-1}c_{m,n}g(l-na)e^{2\pi imbl/L},
\end{equation} for $l\in[L]$.
\end{definition}
Since we will deal with analysis denoising in this paper, we will only focus on $\Phi_g$ from now on.

\subsection{Spark deficient Gabor frames}
Let us first introduce some basic notions needed in this subsection.
\begin{definition}
The symplectic group $\mathrm{SL}(2,\inte)$ consists of all matrices
\begin{equation}
    G=\begin{pmatrix}
\alpha & \beta\\
\gamma & \delta
\end{pmatrix}
\end{equation}
such that $\alpha,\,\beta,\,\gamma,\,\delta\in\inte$ and
\begin{equation}
    \alpha\delta-\beta\gamma\equiv1(\mathrm{mod}L).
\end{equation}
To each such matrix corresponds a unitary matrix given by the explicit formula \cite{dang}
\begin{equation}\label{sic}
    U_G=\frac{e^{i\theta}}{\sqrt{L}}\sum_{u,v=0}^{L-1}\tau^{\beta^{-1}(\alpha v^2-2uv+\delta u^2)}\lvert u\rangle\langle v\rvert,
\end{equation}
where $\theta$ is an arbitrary phase, $\beta^{-1}$ is the inverse\footnote{$\beta\beta^{-1}\equiv1\mathrm{mod}L$} of $\beta\mathrm{mod}L$ and
\begin{equation}
    \tau=-e^{\frac{i\pi}{L}}.
\end{equation}
\end{definition}

\begin{definition}
The \textit{spark} of a set $F$ --denoted by $\mathrm{sp}(F)$-- of $P$ vectors in $\complex$ is the size of the smallest linearly dependent subset of $F$. A frame $F$ is full spark if and only if every set of $L$ elements of $F$ is a basis, or equivalently $\mathrm{sp}(F)=L+1$, otherwise it is spark deficient.
\end{definition}

Based on the previous definition, a Gabor frame with\linebreak $P=L^2/ab$ elements of the form \eqref{gaborsystem} is full spark, if and only if every set of $L$ of its elements is a basis. Now, as proven in \cite{mal}, almost all window vectors generate full spark Gabor frames, so the SDGFs are generated by exceptional window vectors. Indeed, the following theorem was proven in \cite{dang} and informally stated in \cite{sparkmal}, for the \textit{Zauner} matrix $\mathcal{Z}\in\mathrm{SL}(2,\inte)$ given by
\begin{equation}\label{zauner}
    \mathcal{Z}=\begin{pmatrix}
0 & -1\\
1 & -1
\end{pmatrix}\equiv\begin{pmatrix}
0 & L-1\\
1 & L-1
\end{pmatrix}.
\end{equation}
\begin{theorem}[\citep{dang}]\label{sdgf}
Let $L\in\mathbb{Z}$ such that $2\nmid L$, $3\,\vert\,L$ and $L$ is square-free. Then, any eigenvector of the Zauner unitary matrix $U_{\mathcal{Z}}$ (produced by combining \eqref{sic} and \eqref{zauner}), generates a spark deficient Gabor frame for $\complex$.
\end{theorem}

\begin{figure}[b!]
    \centering
    \includegraphics[width=0.5\linewidth]{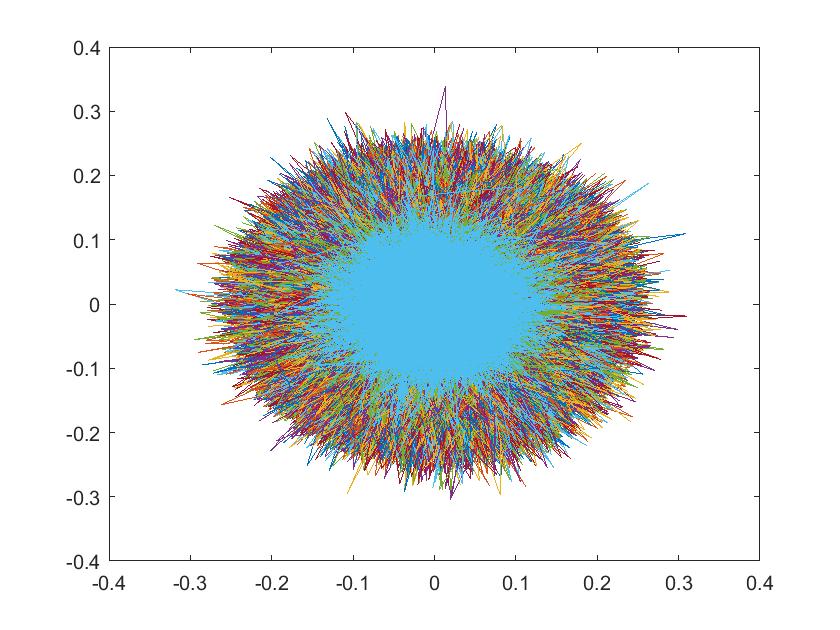}
    \caption{Plotted star-DGT coefficients for the audio signal \textit{Glockenspiel} \cite{ltfat} (262144 samples, $(L,a,b)=(200583,19,23)$)}
    \label{glock}
\end{figure}

Therefore, in order to produce a SDGF and apply its associated analysis operator in \eqref{regl1}, we must first choose an ambient dimension $L$ that fits the assumptions of Theorem \ref{sdgf}. Then, we calculate $U_\mathcal{Z}$ using \eqref{sic} and \eqref{zauner} and in the end, perform its spectral decomposition in order to acquire its eigenvectors. Since all the eigenvectors of $U_\mathcal{Z}$ generate SDGFs, we may choose an arbitrary one, which we call \textit{star window} from now on and denote it as $g_*$. We call the analysis operator associated with such a SDGF \textit{star-DGT} and denote it $\Phi_{g_*}$, in order to indicate the dependance on $g_*$. We coin the term "star", due to the slight resemblance of this DGT to a star when plotted in MATLAB, as it is demonstrated in the example Fig. \ref{glock}.

\begin{remark}\label{lchoice}
A simple way to choose $L$, is by considering its prime factorization: take $k$ prime numbers $p_1^{\alpha_1},\dots,p_k^{\alpha_k}$, with $\alpha_1,\dots,\alpha_k$ not all a multiple of 2 and $p_1=3,p_i\neq2,i=2,\dots,k$, such that $L=3^{\alpha_1}p_2^{\alpha_2}\cdot\dots\cdot p_k^{\alpha_k}$. Since $a,b\,\lvert\,L$, we may also choose $a,b$ to be one, or a multiplication of more than one, prime numbers from the prime factorization of $L$. We have seen empirically that this method for fixing $(L,a,b)$ produces satisfying results, as it is illustrated in the figures of the next pages.
\end{remark}

\section{Experimental Setup}
\subsection{Signals' description and preprocessing}
We run experiments on 12 real-world, real-valued speech signals, all sampled at $16$kHz, taken from \textit{LibriSpeech} corpus \cite{libri}. Their labels along with short description can be found in Table~\ref{sigdec}. The true ambient dimension of each real-world signal does not usually match the conditions of Theorem \ref{sdgf}. Hence, we load each signal and use Remark \ref{lchoice} to cut it off to a specific ambient dimension (from now we shall refer to it as \emph{artificial dimension}) $L$, being as closer as it gets to its true dimension, in order to both denoise a meaningful part of the signal and meet the conditions of Theorem \ref{sdgf}.

\begin{table}[h]
    \caption{Signals' details}
    \centering
    \scalebox{0.8}{\begin{tabular}{|| c | c | c | c | c ||}
    \hline
     $\#$ & Labels & True ambient dimension & Artificial dimension $L$ & Types of noise added\\
     \hline\hline
     1 & 251-136532-0014 & 36240 & $33915$ & Gaussian and pink\\
     \hline
     2 & 8842-304647-0007 & 27680 & $27531$ & Gaussian and blue\\
     \hline
     3 & 2035-147960-0013 & 42800 & $41769$ & Gaussian and pink\\
     \hline
     4 & 1462-170145-0020 & 34400 & $33915$ & Gaussian and blue\\
     \hline
     5 & 6241-61943-0002 & 43760 & $43605$ & Gaussian and blue\\
     \hline
     6 & 5338-284437-0025 & 31040 & $29835$ & Gaussian and blue\\
     \hline
     7 & 3752-4944-0042 & 51360 & $51051$ & Gaussian and blue\\
     \hline
     8 & 5694-64038-0013 & 52880 & $51051$ & Gaussian and pink\\
     \hline
     9 & 5895-34615-0001 & 52880 & $51051$ & Gaussian and pink\\
     \hline
     10 & 2428-83699-0035 & 43600 & $41769$ & Gaussian and pink\\
     \hline
     11 & 2803-154320-0006 & 34880 & $33915$ & Gaussian and blue\\
     \hline
     12 & 3752-4944-0008 & 31040 & $29835$ & Gaussian and pink\\
     \hline
    \end{tabular}}
    \label{sigdec}
\end{table}

\begin{table}[]
    \centering
    \caption{251-136532-0014 with $L=33915$ and $\sigma=0.001$}
    \begin{subtable}[h]{0.90\textwidth}
    \scalebox{0.75}{
    \begin{tabular}{|| c | c | c | c | c | c ||}
    \hline
    & \multicolumn{5}{|c|}{MSEs}\\
    \hline
    \diagbox{Windows}{$(a,b)$}
    & $(15,15)$ & $(5,17)$ & $(7,19)$ & $(17,19)$ & $(21,21)$\\
    \hline
    Gaussian & $ 8.4929\cdot10^{-4}$ & $8.4846\cdot10^{-4}$ &  $8.4233\cdot10^{-4}$ & $8.4488\cdot10^{-4}$ & $8.4188\cdot10^{-4}$\\
    \hline
    Hann & $ 8.4917\cdot10^{-4}$ & $8.4858\cdot10^{-4}$ & $ 8.4222\cdot10^{-4}$ & $8.4502\cdot10^{-4}$ & $8.4196\cdot10^{-4}$\\
    \hline
    Hamming & $ 8.4936\cdot10^{-4}$ & $8.4830\cdot10^{-4}$ & $8.4234\cdot10^{-4}$ & $8.4490\cdot10^{-4}$ & $8.4199\cdot10^{-4}$\\
    \hline
    Star & $\bf7.4783\cdot10^{-4}$ & $\bf8.1694\cdot10^{-4}$ & $\bf7.5914\cdot10^{-4}$ & $\bf\color{RedViolet}6.7464\cdot10^{-4}$ & $\bf7.0497\cdot10^{-4}$\\
    \hline
    \end{tabular}}
    \captionsetup{justification=centering}
    \caption{Gaussian noise}
    \label{gaussianpar2}
    \end{subtable}\vfill
    \begin{subtable}[h]{0.9\textwidth}
    \scalebox{0.75}{
    \begin{tabular}{|| c | c | c | c | c | c ||}
    \hline
    & \multicolumn{5}{|c|}{MSEs}\\
    \hline
    \diagbox{Windows}{$(a,b)$}
    & $(15,15)$ & $(5,17)$ & $(7,19)$ & $(17,19)$ & $(21,21)$\\
    \hline
    Gaussian & $  5.3213\cdot10^{-4}$ & $5.3660\cdot10^{-4}$ &  $5.3533\cdot10^{-4}$ & $5.3625\cdot10^{-4}$ & $5.4130\cdot10^{-4}$\\
    \hline
    Hann & $5.3217\cdot10^{-4}$ & $5.3655\cdot10^{-4}$ & $5.3521\cdot10^{-4}$ & $5.3620\cdot10^{-4}$ & $5.4122\cdot10^{-4}$\\
    \hline
    Hamming & $5.3213\cdot10^{-4}$ & $5.3655\cdot10^{-4}$ & $5.3528\cdot10^{-4}$ & $5.3632\cdot10^{-4}$ & $5.4111\cdot10^{-4}$\\
    \hline
    Star & $\bf4.5283\cdot10^{-4}$ & $\bf4.7100\cdot10^{-4}$ & $\bf4.6333\cdot10^{-4}$ & $\bf\color{RedViolet}3.5476\cdot10^{-4}$ & $\bf4.2386\cdot10^{-4}$\\
    \hline
    \end{tabular}}
    \captionsetup{justification=centering}
    \caption{Pink noise}
    \label{pinkpar2}
    \end{subtable}
    \label{251par}
\end{table}

\begin{table}[]
\centering
    \caption{2035-147960-0013 with $L=41769$ and $\sigma=0.001$}
    \begin{subtable}[h]{0.9\textwidth}
    \scalebox{0.75}{
    \begin{tabular}{|| c | c | c | c | c | c ||}
    \hline
    & \multicolumn{5}{|c|}{MSEs}\\
    \hline
    \diagbox{Windows}{$(a,b)$} & $(9,9)$ & $(7,13)$ & $(9,17)$ & $(13,17)$ & $(17,17)$\\
    \hline
    Gaussian & $ 9.1225\cdot10^{-4}$ & $9.1113\cdot10^{-4}$ & $9.1134\cdot10^{-4}$ & $9.0846\cdot10^{-4}$ & $9.1484\cdot10^{-4}$\\
    \hline
    Hann & $ 9.1230\cdot10^{-4}$ & $9.1111\cdot10^{-4}$ & $ 9.1126\cdot10^{-4}$ & $9.0833\cdot10^{-4}$ & $9.1492\cdot10^{-4}$\\
    \hline
    Hamming & $ 9.1221\cdot10^{-4}$ & $9.1108\cdot10^{-4}$ & $9.1130\cdot10^{-4}$ & $9.0838\cdot10^{-4}$ & $9.1492\cdot10^{-4}$\\
    \hline
    Star & $\bf8.5635\cdot10^{-4}$ & $\bf8.0955\cdot10^{-4}$ & $\bf8.2828\cdot10^{-4}$ & $\bf\color{RedViolet}7.7499\cdot10^{-4}$ & $\bf8.1464\cdot10^{-4}$\\
    \hline
    \end{tabular}}
    \captionsetup{justification=centering}
    \caption{Gaussian noise}
    \label{gaussianpar1}
    \end{subtable}\vfill
    \begin{subtable}[h]{0.9\textwidth}
    \scalebox{0.75}{
    \begin{tabular}{|| c | c | c | c | c | c ||}
    \hline
    & \multicolumn{5}{|c|}{MSEs}\\
    \hline
    \diagbox{Windows}{$(a,b)$} & $(9,9)$ & $(7,13)$ & $(9,17)$ & $(13,17)$ & $(17,17)$\\
    \hline
    Gaussian & $5.8699\cdot10^{-4}$ & $ 5.8925\cdot10^{-4}$ & $ 5.8942\cdot10^{-4}$ & $ 5.9071\cdot10^{-4}$ & $5.8948\cdot10^{-4}$\\
    \hline
    Hann & $5.8710\cdot10^{-4}$ & $5.8921\cdot10^{-4}$ & $ 5.8945\cdot10^{-4}$ & $ 5.9080\cdot10^{-4}$ & $5.8949\cdot10^{-4}$\\
    \hline
    Hamming & $5.8703\cdot10^{-4}$ & $ 5.8928\cdot10^{-4}$ & $ 5.8961\cdot10^{-4}$ & $ 5.9060\cdot10^{-4}$ & $5.8943\cdot10^{-4}$\\
    \hline
    Star & $\bf5.3307\cdot10^{-4}$ & $\bf5.3072\cdot10^{-4}$ & $\bf4.9653\cdot10^{-4}$ & $\bf\color{RedViolet}4.8565\cdot10^{-4}$ & $\bf4.8874\cdot10^{-4}$\\
    \hline
    \end{tabular}}
    \captionsetup{justification=centering}
    \caption{Pink noise}
    \label{pinkpar3}
    \end{subtable}
    \label{2035par}
\end{table}

\begin{table}[]
    \caption{5694-64038-0013 with $\sigma=0.001$ and Gaussian noise}
    \centering
    \begin{subtable}[h]{0.9\textwidth}
    \scalebox{0.75}{\begin{tabular}{|| c | c | c | c | c | c ||}
    \hline
    & \multicolumn{5}{|c|}{MSEs}\\
    \hline
    \diagbox{Windows}{$(a,b)$} & $(11,11)$ & $(13,13)$
    & $(13,21)$ & $(11,17)$ & $(13,17)$\\
    \hline
    Gaussian & $8.7630\cdot10^{-4}$ & $8.7634\cdot10^{-4}$ & $8.7914\cdot10^{-4}$ & $8.7299\cdot10^{-4}$ & $ 8.7429\cdot10^{-4}$\\
    \hline
    Hann & $8.7643\cdot10^{-4}$ & $8.7641\cdot10^{-4}$ & $8.7918\cdot10^{-4}$ & $ 8.7300\cdot10^{-4}$ & $ 8.7419\cdot10^{-4}$\\
    \hline
    Hamming & $8.7629\cdot10^{-4}$ & $8.7639\cdot10^{-4}$ & $8.7909\cdot10^{-4}$ & $8.7301\cdot10^{-4}$ & $ 8.7431\cdot10^{-4}$\\
    \hline
    Star & $\bf 8.2926\cdot10^{-4}$ & $\bf 8.0809\cdot10^{-4}$ & $\bf 7.6079\cdot10^{-4}$ & $\bf 7.5266\cdot10^{-4}$ & $\bf\color{RedViolet}7.5009\cdot10^{-4}$ \\
    \hline
    \end{tabular}
    }
    \captionsetup{justification=centering}
    \caption{Gaussian noise}
    \label{gaussianpar5}
    \end{subtable}
    \begin{subtable}[h]{0.9\textwidth}
    \scalebox{0.75}{\begin{tabular}{|| c | c | c | c | c | c ||}
    \hline
    & \multicolumn{5}{|c|}{MSEs}\\
    \hline
    \diagbox{Windows}{$(a,b)$} & $(11,11)$ & $(13,13)$
    & $(13,21)$ & $(11,17)$ & $(13,17)$\\
    \hline
    Gaussian & $5.5950\cdot10^{-4}$ & $5.5945\cdot10^{-4}$ & $5.6133\cdot10^{-4}$ & $5.5808\cdot10^{-4}$ & $5.5888\cdot10^{-4}$\\
    \hline
    Hann & $5.5949\cdot10^{-4}$ & $5.5943\cdot10^{-4}$ & $5.6131\cdot10^{-4}$ & $5.5812\cdot10^{-4}$ & $5.5890\cdot10^{-4}$\\
    \hline
    Hamming & $5.5949\cdot10^{-4}$ & $5.5958\cdot10^{-4}$ & $5.6128\cdot10^{-4}$ & $5.5800\cdot10^{-4}$ & $5.5897\cdot10^{-4}$\\
    \hline
    Star & $\bf 5.1971\cdot10^{-4}$ & $\bf 4.9725\cdot10^{-4}$ & $\bf 4.4889\cdot10^{-4}$ & $\bf 4.7244\cdot10^{-4}$ & $\bf\color{RedViolet}4.3773\cdot10^{-4}$ \\
    \hline
    \end{tabular}
    }
    \captionsetup{justification=centering}
    \caption{Pink noise}
    \label{pinkpar5}
    \end{subtable}
    \label{5694par}
\end{table}

\begin{table}[]
    \caption{5338-284437-0025 with $L=29835$ and $\sigma=0.001$}
    \centering
    \begin{subtable}[h]{0.8\textwidth}
    \scalebox{0.75}{\begin{tabular}{|| c  | c | c | c | c | c ||}
    \hline
    & \multicolumn{5}{|c|}{MSEs}\\
    \hline
    \diagbox{Windows}{$(a,b)$}
    & $(9,9)$ & $(15,15)$ & $(5,13)$ & $(5,17)$ & $(13,17)$\\
    \hline
    Gaussian & $12\cdot10^{-4}$ & $12\cdot10^{-4}$ & $12\cdot10^{-4}$ & $12\cdot10^{-4}$ & $12\cdot10^{-4}$\\
    \hline
    Hann & $12\cdot10^{-4}$ & $12\cdot10^{-4}$ & $12\cdot10^{-4}$ & $12\cdot10^{-4}$ & $12\cdot10^{-4}$\\
    \hline
    Hamming & $12\cdot10^{-4}$ & $12\cdot10^{-4}$ & $12\cdot10^{-4}$ & $12\cdot10^{-4}$ & $12\cdot10^{-4}$\\
    \hline
    Star & $\bf11\cdot10^{-4}$ & $\bf9.8065\cdot10^{-4}$ & $\bf11\cdot10^{-4}$ & $\bf11\cdot10^{-4}$ & $\bf\color{RedViolet}9.7897\cdot10^{-4}$\\
    \hline
    \end{tabular}
    }
    \captionsetup{justification=centering}
    \caption{Gaussian noise}
    \label{gaussianpar4}
    \end{subtable}
    \begin{subtable}[h]{0.9\textwidth}
    \scalebox{0.75}{\begin{tabular}{|| c  | c | c | c | c | c ||}
    \hline
    & \multicolumn{5}{|c|}{MSEs}\\
    \hline
    \diagbox{Windows}{$(a,b)$}
    & $(9,9)$ & $(15,15)$ & $(5,13)$ & $(5,17)$ & $(13,17)$\\
    \hline
    Gaussian & $7.8626\cdot10^{-4}$ & $7.8715\cdot10^{-4}$ & $7.8627\cdot10^{-4}$ & $7.8690\cdot10^{-4}$ & $7.8674\cdot10^{-4}$\\
    \hline
    Hann & $7.8629\cdot10^{-4}$ & $7.8689\cdot10^{-4}$ & $7.8650\cdot10^{-4}$ & $7.8689\cdot10^{-4}$ & $7.8696\cdot10^{-4}$\\
    \hline
    Hamming & $7.8627\cdot10^{-4}$ & $7.8745\cdot10^{-4}$ & $7.8623\cdot10^{-4}$ & $7.8676\cdot10^{-4}$ & $7.8677\cdot10^{-4}$\\
    \hline
    Star & $\bf6.8879\cdot10^{-4}$ & $\bf6.6695\cdot10^{-4}$ & $\bf7.0199\cdot10^{-4}$ & $\bf6.6884\cdot10^{-4}$ & $\bf\color{RedViolet}6.3867\cdot10^{-4}$\\
    \hline
    \end{tabular}
    }
    \captionsetup{justification=centering}
    \caption{Blue noise}
    \label{bluepar4}
    \end{subtable}
    \label{5338par}
\end{table}

\begin{table}[]
    \caption{6241-61943-0002 with $L=43605$ and $\sigma=0.001$}
    \centering
    \begin{subtable}[h]{0.9\textwidth}
    \scalebox{0.75}{\begin{tabular}{|| c  | c | c | c | c | c ||}
    \hline
    & \multicolumn{5}{|c|}{MSEs}\\
    \hline
    \diagbox{Windows}{$(a,b)$}
    & $(9,9)$ & $(5,17)$ & $(9,17)$ & $(17,19)$ & $(19,19)$\\
    \hline
    Gaussian & $5.0929\cdot10^{-4}$ & $5.1122\cdot10^{-4}$ & $5.1164\cdot10^{-4}$ & $5.0961\cdot10^{-4}$ & $ 5.1202\cdot10^{-4}$\\
    \hline
    Hann & $5.0929\cdot10^{-4}$ & $5.1119\cdot10^{-4}$ & $ 5.1166\cdot10^{-4}$ & $5.0961\cdot10^{-4}$ & $ 5.1202\cdot10^{-4}$\\
    \hline
    Hamming & $5.0928\cdot10^{-4}$ & $5.1121\cdot10^{-4}$ & $5.1160\cdot10^{-4}$ & $5.0957\cdot10^{-4}$ & $ 5.1201\cdot10^{-4}$\\
    \hline
    Star & $\bf4.9231\cdot10^{-4}$ & $\bf4.6263\cdot10^{-4}$ & $\bf4.7426\cdot10^{-4}$ & $ \bf\color{RedViolet}4.5764\cdot10^{-4}$ & $\bf4.6173\cdot10^{-4}$\\
    \hline
    \end{tabular}
    }
    \captionsetup{justification=centering}
    \caption{Gaussian noise}
    \label{gaussianpar3}
    \end{subtable}
    \begin{subtable}[h]{0.9\textwidth}
    \scalebox{0.75}{\begin{tabular}{|| c  | c | c | c | c | c ||}
    \hline
    & \multicolumn{5}{|c|}{MSEs}\\
    \hline
    \diagbox{Windows}{$(a,b)$}
    & $(9,9)$ & $(5,17)$ & $(9,17)$ & $(17,19)$ & $(19,19)$\\
    \hline
    Gaussian & $2.6371\cdot10^{-4}$ & $2.6358\cdot10^{-4}$ & $2.6336\cdot10^{-4}$ & $2.6356\cdot10^{-4}$ & $2.6356\cdot10^{-4}$\\
    \hline
    Hann & $2.6373\cdot10^{-4}$ & $2.6360\cdot10^{-4}$ & $2.6338\cdot10^{-4}$ & $2.6359\cdot10^{-4}$ & $2.6359\cdot10^{-4}$\\
    \hline
    Hamming & $2.6371\cdot10^{-4}$ & $2.6359\cdot10^{-4}$ & $2.6340\cdot10^{-4}$ & $2.6359\cdot10^{-4}$ & $2.6359\cdot10^{-4}$\\
    \hline
    Star & $\bf2.3890\cdot10^{-4}$ & $\bf2.3582\cdot10^{-4}$ & $\bf2.2962\cdot10^{-4}$ & $ \bf\color{RedViolet}1.9016\cdot10^{-4}$ & $\bf2.2279\cdot10^{-4}$\\
    \hline
    \end{tabular}
    }
    \captionsetup{justification=centering}
    \caption{Blue noise}
    \label{bluepar3}
    \end{subtable}
    \label{6241par}
\end{table}

\begin{table}[]
    \caption{8842-304647-0007 with $L=27531$ and $\sigma=0.001$}
    \centering
    \begin{subtable}[h]{0.9\textwidth}
    \scalebox{0.75}{\begin{tabular}{|| c | c | c | c | c | c ||}
    \hline
    & \multicolumn{5}{|c|}{MSEs}\\
    \hline
    \diagbox{Windows}{$(a,b)$} & $(9,9)$
    & $(9,23)$ & $(7,19)$ & $(19,19)$ & $(19,23)$\\
    \hline
    Gaussian & $7.2315\cdot10^{-4}$ & $ 7.2621\cdot10^{-4}$ & $7.1806\cdot10^{-4}$ & $7.2532\cdot10^{-4}$ & $ 7.1924\cdot10^{-4}$\\
    \hline
    Hann & $7.2314\cdot10^{-4}$ & $ 7.2605\cdot10^{-4}$ & $7.1806\cdot10^{-4}$ & $7.2535\cdot10^{-4}$ & $ 7.1915\cdot10^{-4}$\\
    \hline
    Hamming & $7.2321\cdot10^{-4}$ & $ 7.2609\cdot10^{-4}$ & $7.1803\cdot10^{-4}$ & $7.2519\cdot10^{-4}$ & $ 7.1919\cdot10^{-4}$\\
    \hline
    Star & $\bf6.7449\cdot10^{-4}$ & $\bf6.1704\cdot10^{-4}$ & $\bf6.0620\cdot10^{-4}$ & $\bf6.2294\cdot10^{-4}$ & $\bf\color{RedViolet}5.4646\cdot10^{-4}$ \\
    \hline
    \end{tabular}}
    \captionsetup{justification=centering}
    \caption{Gaussian noise}
    \label{gaussianpar6}
    \end{subtable}
    \begin{subtable}[h]{0.9\textwidth}
    \scalebox{0.75}{\begin{tabular}{|| c | c | c | c | c | c ||}
    \hline
    & \multicolumn{5}{|c|}{MSEs}\\
    \hline
    \diagbox{Windows}{$(a,b)$} & $(9,9)$
    & $(9,23)$ & $(7,19)$ & $(19,19)$ & $(19,23)$\\
    \hline
    Gaussian & $4.1860\cdot10^{-4}$ & $4.1906\cdot10^{-4}$ & $4.1881\cdot10^{-4}$ & $4.1925\cdot10^{-4}$ & $4.1887\cdot10^{-4}$\\
    \hline
    Hann & $4.1863\cdot10^{-4}$ & $4.1903\cdot10^{-4}$ & $4.1879\cdot10^{-4}$ & $4.1935\cdot10^{-4}$ & $4.1887\cdot10^{-4}$\\
    \hline
    Hamming & $4.1864\cdot10^{-4}$ & $4.1899\cdot10^{-4}$ & $4.1885\cdot10^{-4}$ & $4.1932\cdot10^{-4}$ & $4.1884\cdot10^{-4}$\\
    \hline
    Star & $\bf3.6969\cdot10^{-4}$ & $\bf2.8825\cdot10^{-4}$ & $\bf3.3767\cdot10^{-4}$ & $\bf3.2135\cdot10^{-4}$ & $\bf\color{RedViolet}2.8677\cdot10^{-4}$ \\
    \hline
    \end{tabular}}
    \captionsetup{justification=centering}
    \caption{Blue noise}
    \label{bluepar6}
    \end{subtable}
    \label{8842par}
\end{table}

\subsection{Experimental settings}
\begin{enumerate}
    \item We examine different pairs of time-frequency parameters $(a,b)$, deploying Remark \ref{lchoice}.
    \item We use the power iteration method \cite{power} which yields the largest in magnitude eigenvalue and  corresponding eigenvector of  $U_{\mathcal{Z}}$, then set this eigenvector as our desired window vector $g_*$.
    \item We construct --using the MATLAB package \textit{LTFAT} \cite{ltfat}-- four different Gabor frames with their associated analysis operators/DGTs, which go as follows: $\Phi_{g_1}$, $\Phi_{g_2}$, $\Phi_{g_3}$ and  $\Phi_{g_*}$, corresponding to a Gaussian, a Hann, a Hamming and the star window vector respectively. Since we process real-valued signals, we alter the four analysis operators to compute only the DGT coefficients of positive frequencies instead of the full DGT coefficients. 
    \item We consider zero-mean Gaussian noise with standard deviation $\sigma$ either varying, i.e. $\sigma\equiv\bm{\sigma}=\text{linspace}(0.001,0.01,100)$ (MATLAB function \emph{linspace} generates a row vector of 100 evenly spaced points between 0.001 and 0.01), or fixed scalar $\sigma=0.001$. Similarly, we consider pink and blue noises with amplitude either varying and controlled by a scaling factor $\bm{\sigma}=\text{linspace}(0.001,0.01,100)$, or fixed and controlled by the scalar $\sigma=0.001$. Note that for $\bm{\sigma_{100}}=0.01$, the amplitude of the coloured\footnote{in the rest of the paper, when we speak of coloured noises, we mean the examined cases of pink and blue noise} noises is almost equal to the amplitude of the signals. 
    \item We take noisy measurements
    \begin{equation}
        y=x+e,
    \end{equation}
    where $e$ denotes either Gaussian or coloured noise.
    \item We solve --using the Matlab package \textit{TFOCS} \cite{tfocs}-- four different instances of \eqref{regl1}, one for each of the four DGTs. For TFOCS, we set $x_0=0,\,z_0=[\,]$; for each of the instances $i=1,2,3,*$, we set the smoothing parameter $\mu_i=10^{-1}\|\Phi_ix\|_\infty$, since we noticed an improved performance of the solving algorithm when $\mu$ is a function of $\Phi_i$ (the scaling factor $10^{-1}$ and the function $\|\cdot\|_\infty$ are simply chosen empirically) and employ the \textit{solver\_BPDN\_W} solver.
    \item From the aforementioned procedure, we obtain four different estimators for $x$, namely $\hat{x}_1$, $\hat{x}_2$, $\hat{x}_3$,  $\hat{x}_*$ and their corresponding MSEs, i.e.\linebreak $\mathbb{E}(\|x-\hat{x}_i\|_2^2)$, $i=1,2,3,*$.
\end{enumerate}

\begin{figure}
     \centering
     \begin{subfigure}[b]{0.4\textwidth}
         \centering         \includegraphics[width=\textwidth]{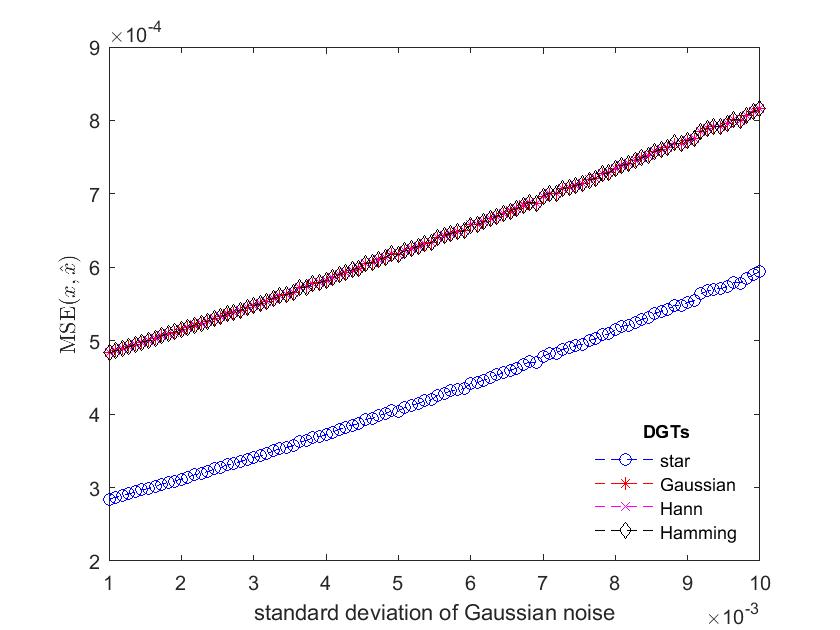}
         \caption{251-136532-0014, $(L,a,b)=(33915,51,19)$, Gaussian noise}
         \label{251}
     \end{subfigure}
     \hfill
     \begin{subfigure}[b]{0.4\textwidth}
         \centering
         \includegraphics[width=\textwidth]{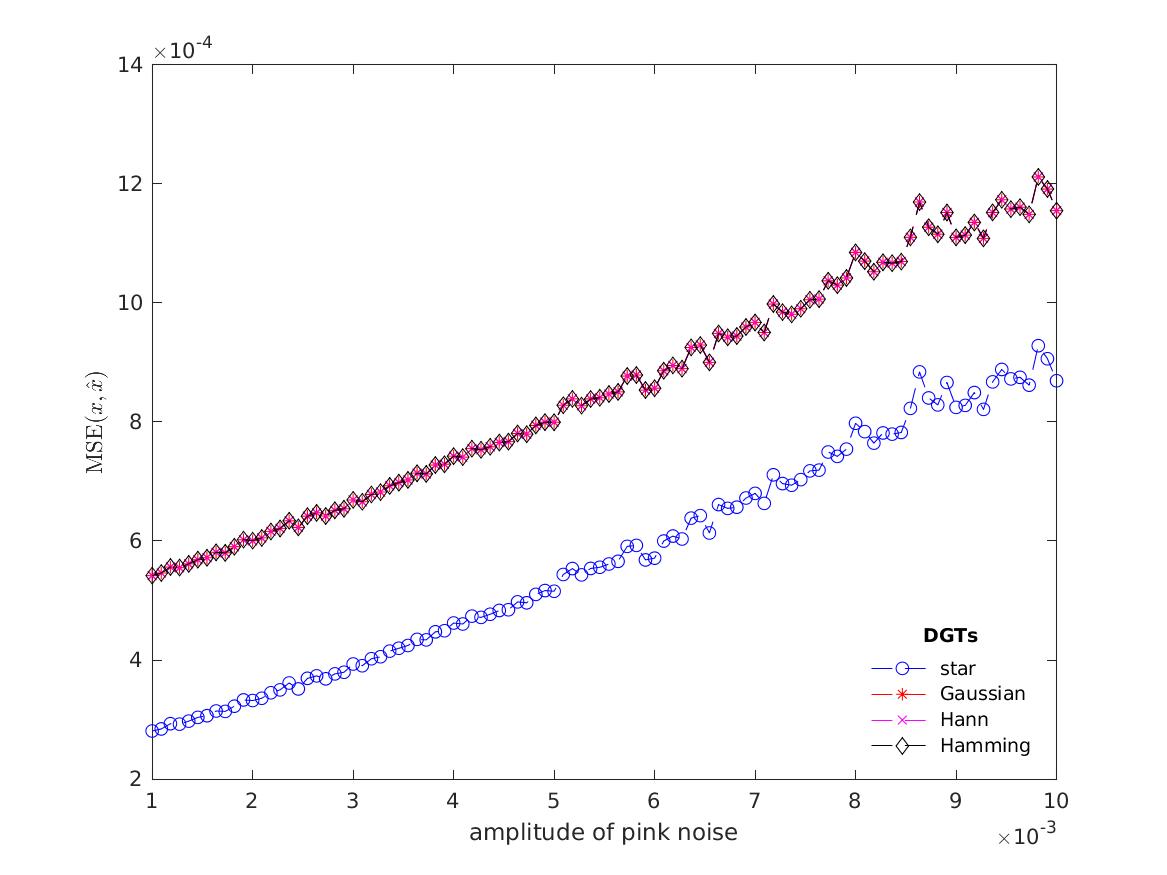}
         \caption{251-136532-0014, $(L,a,b)=(33915,51,19)$, pink noise}
         \label{251pink}
     \end{subfigure}
     \hfill
     \begin{subfigure}[b]{0.4\textwidth}
         \centering
         \includegraphics[width=\textwidth]{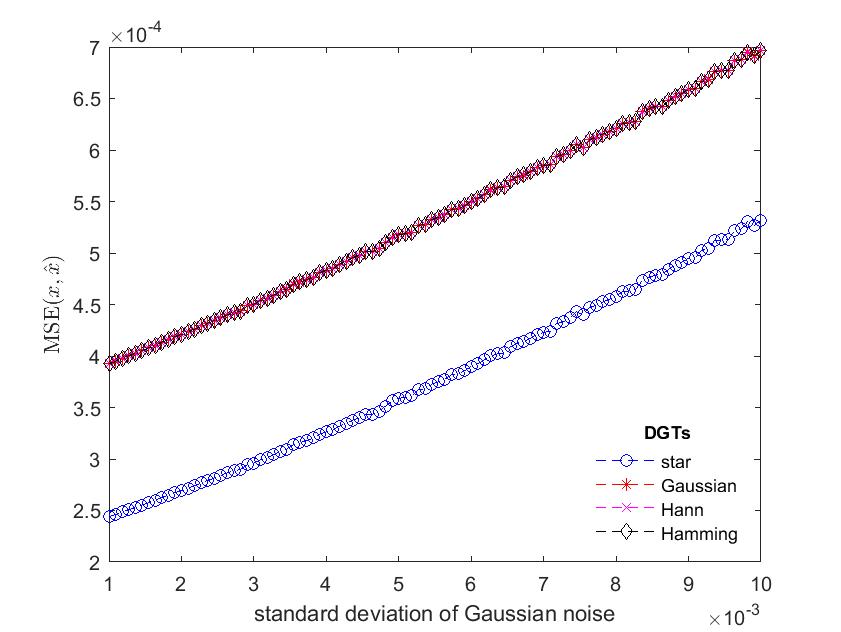}
         \caption{8842-304647-0007, $(L,a,b)=(27531,19,23)$, Gaussian noise}
         \label{8842}
     \end{subfigure}\hfill
     \begin{subfigure}[b]{0.4\textwidth}
         \centering
         \includegraphics[width=\textwidth]{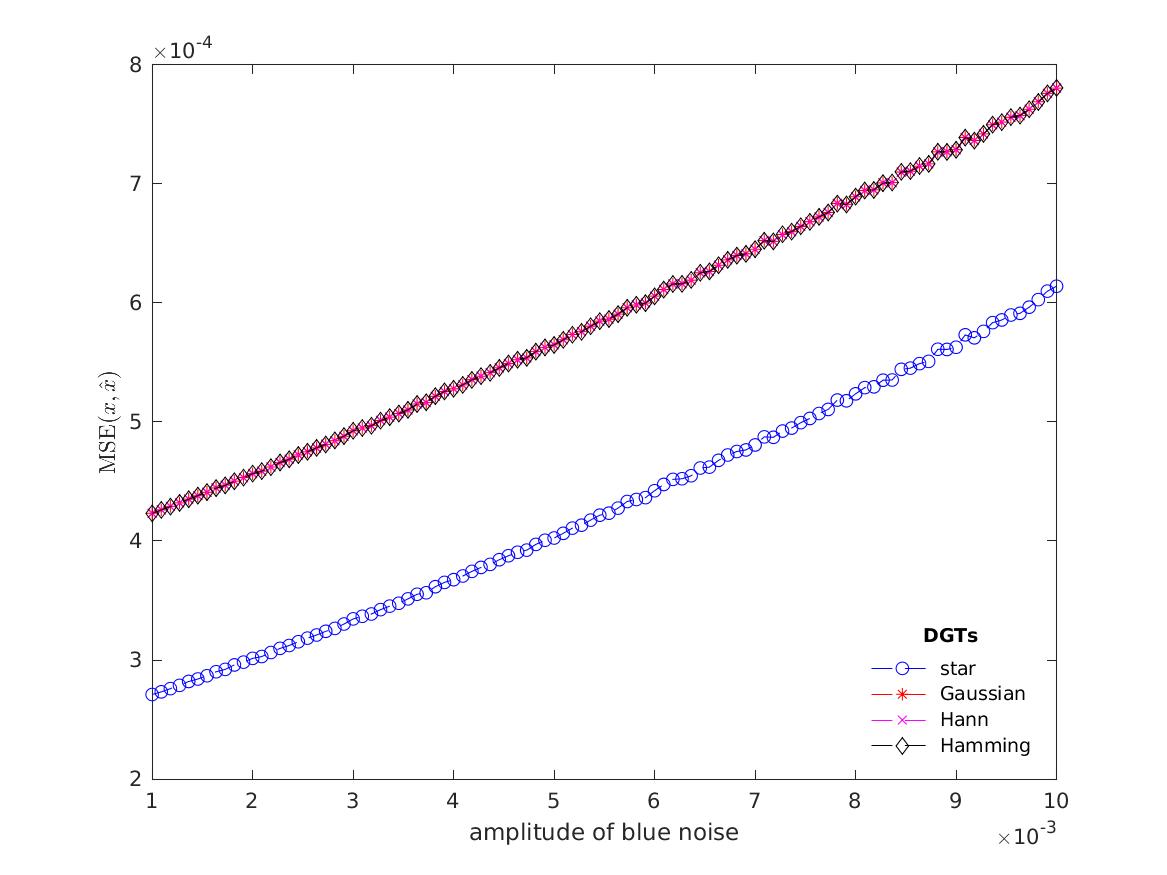}
         \caption{8842-304647-0007, $(L,a,b)=(27531,19,23)$, blue noise}
         \label{8842blue}
     \end{subfigure}
     \hfill
     \begin{subfigure}[b]{0.4\textwidth}
         \centering
         \includegraphics[width=\textwidth]{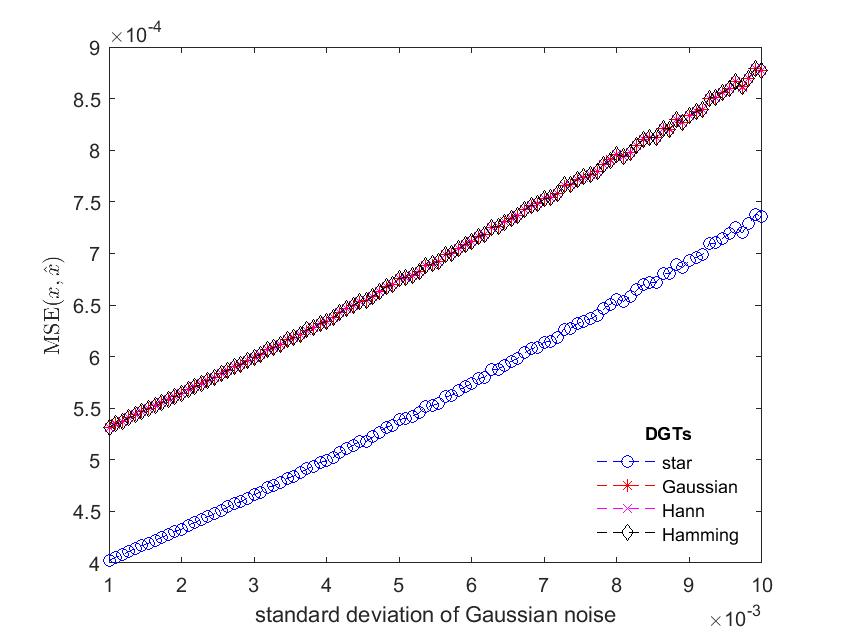}
         \caption{2035-147960-0013, $(L,a,b)=(41769,21, 17)$, Gaussian noise}
         \label{2035}
     \end{subfigure}
     \hfill
     \begin{subfigure}[b]{0.4\textwidth}
         \centering
         \includegraphics[width=\textwidth]{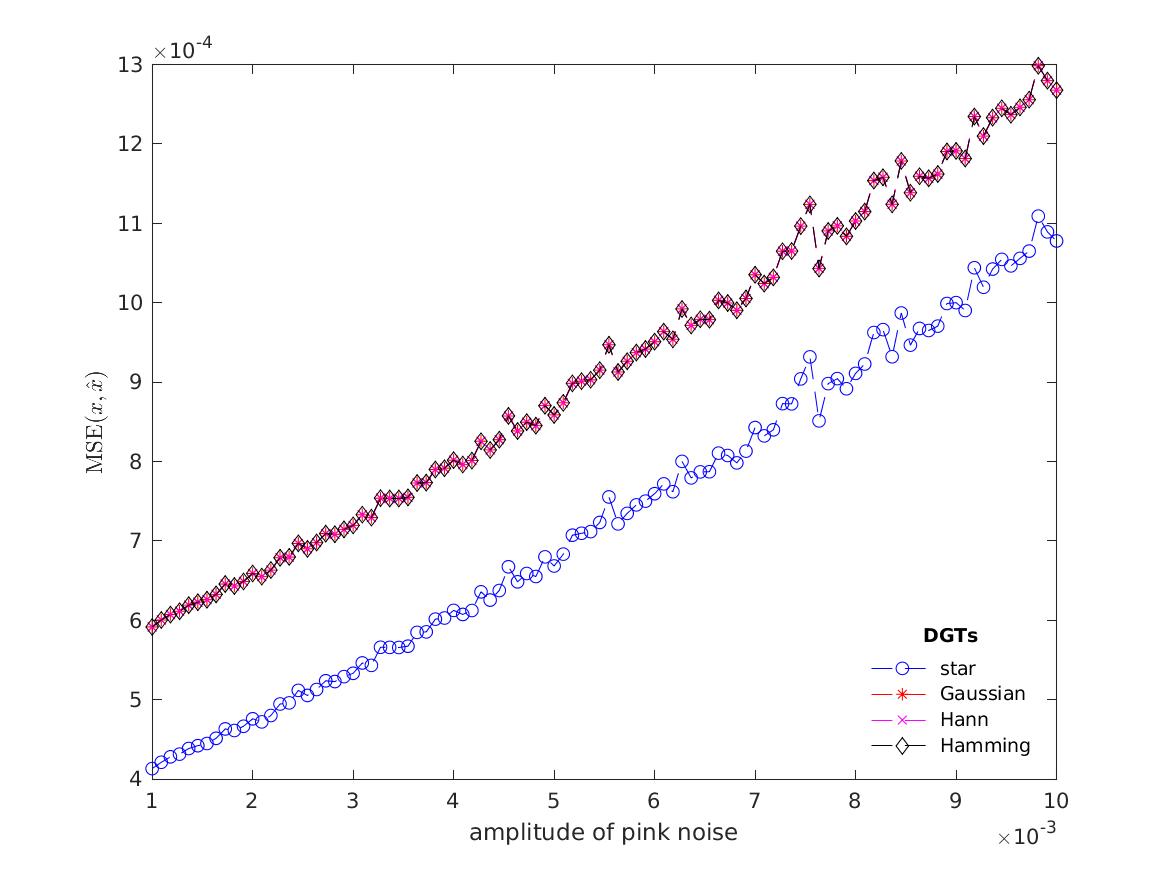}
         \caption{2035-147960-0013, $(L,a,b)=(41769,21, 17)$, pink noise}
         \label{2035pink}
     \end{subfigure}
     \begin{subfigure}[b]{0.4\textwidth}
         \centering
         \includegraphics[width=\textwidth]{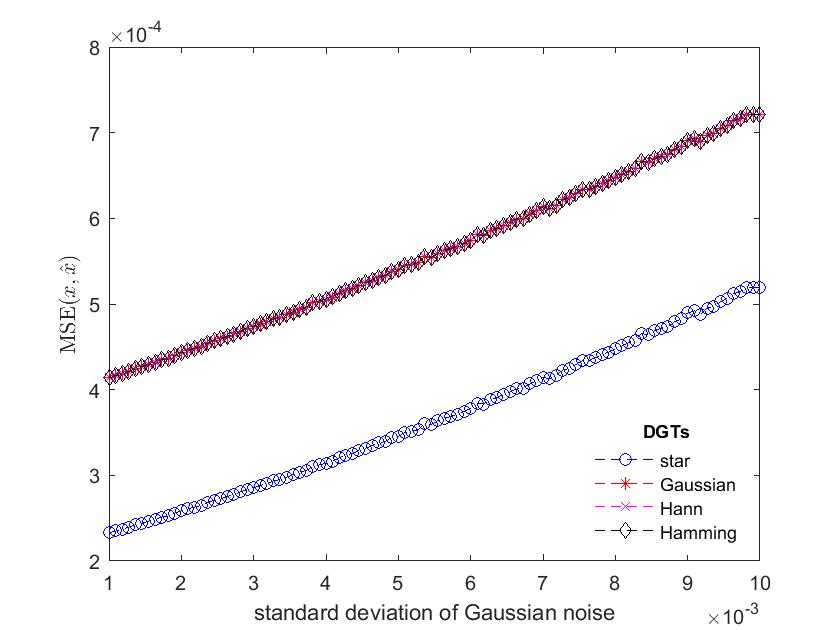}
         \caption{1462-170145-0020, $(L,a,b)=(33915,51, 19)$, Gaussian noise}
         \label{1462}
     \end{subfigure}\hfill
     \begin{subfigure}[b]{0.4\textwidth}
         \centering
         \includegraphics[width=\textwidth]{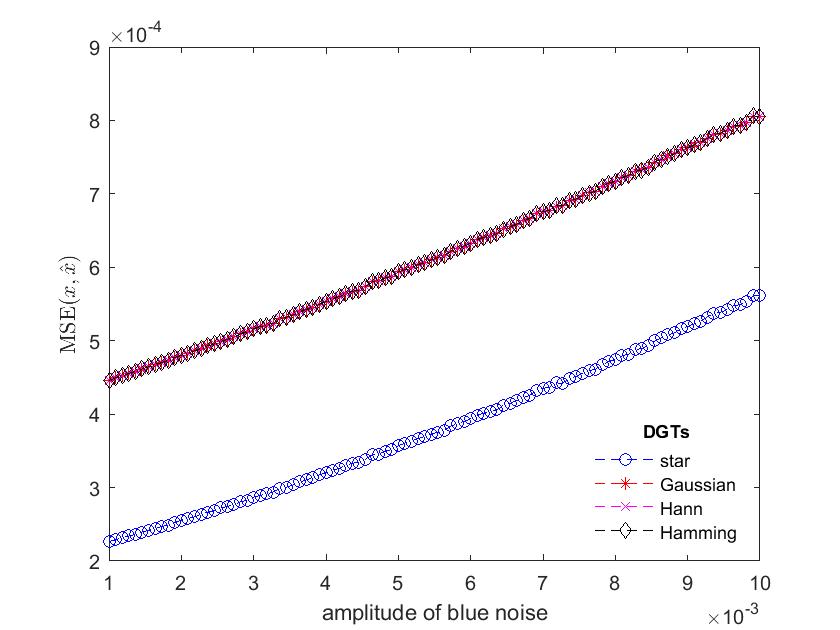}
         \caption{1462-170145-0020, $(L,a,b)=(33915,51, 19)$, blue noise}
         \label{1462blue}
     \end{subfigure}
     \captionsetup{justification=centering}
        \caption{Rate of denoising success for 4 speech signals with different parameters $(L,a,b)$, contaminated by Gaussian (left) and coloured (right) noise. Note that 3 of the 4 methods roughly coincide.}
        \label{gaussian1}
\end{figure}

\begin{figure}
     \centering
     \begin{subfigure}[b]{0.4\textwidth}
         \centering         \includegraphics[width=\textwidth]{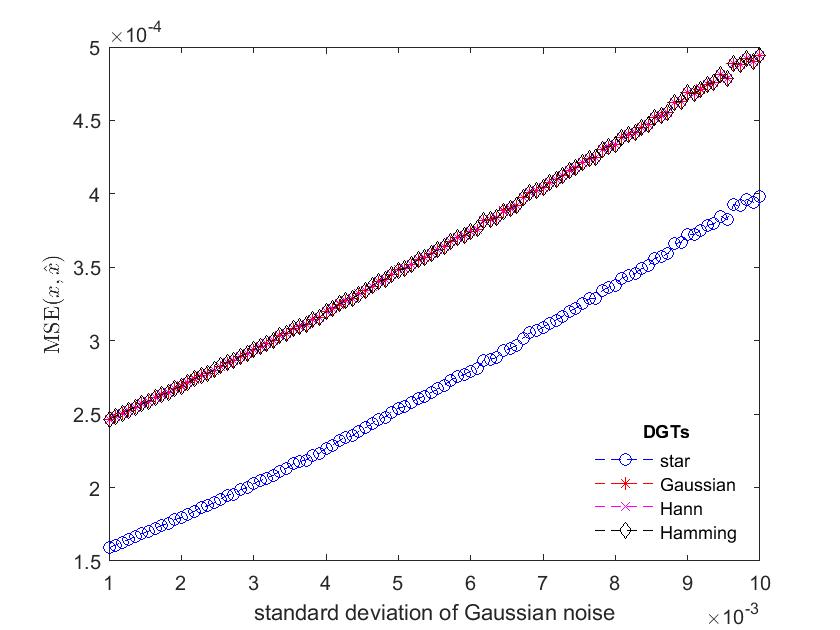}
         \caption{6241-61943-0002, $(L,a,b)=(43605,51,19)$, Gaussian noise}
         \label{6241}
     \end{subfigure}
     \hfill
     \begin{subfigure}[b]{0.4\textwidth}
         \centering
         \includegraphics[width=\textwidth]{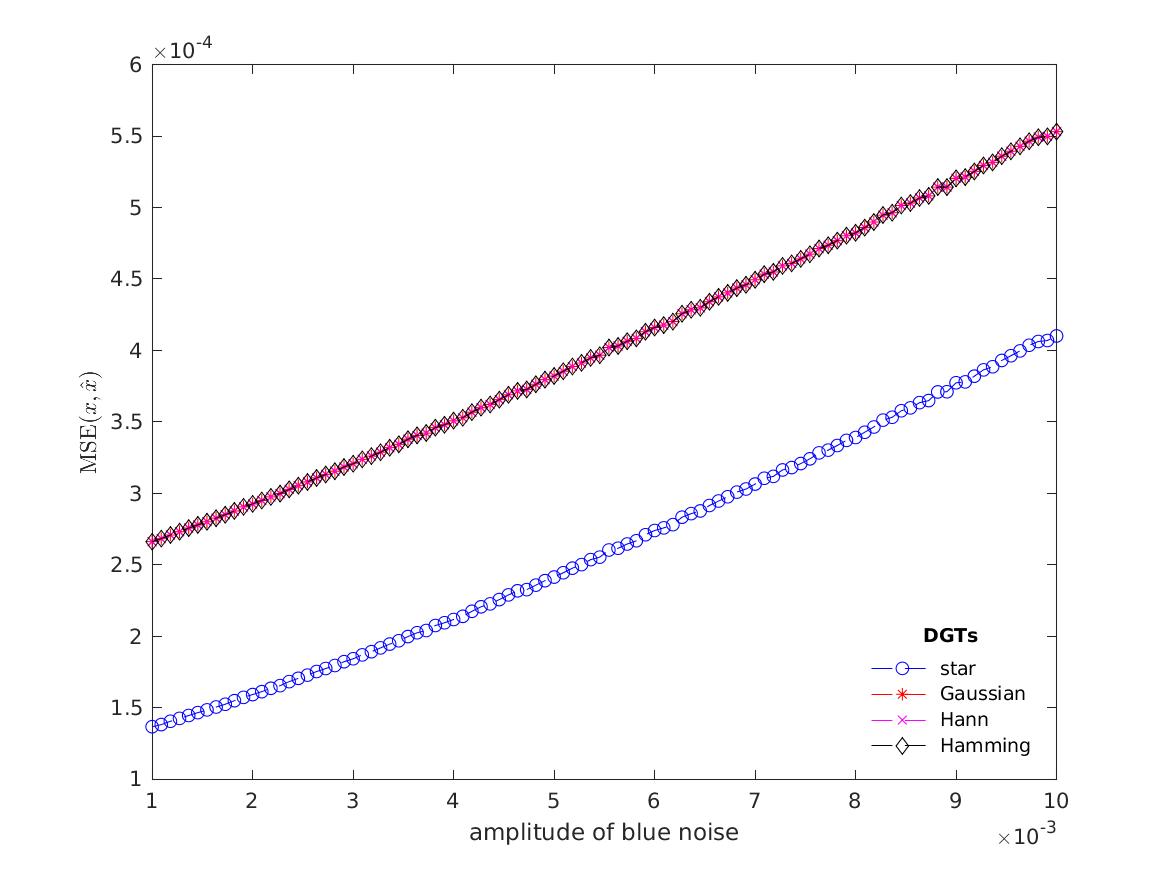}
         \caption{6241-61943-0002, $(L,a,b)=(43605,51,19)$, pink noise}
         \label{6241blue}
     \end{subfigure}
     \hfill
     \begin{subfigure}[b]{0.4\textwidth}
         \centering
         \includegraphics[width=\textwidth]{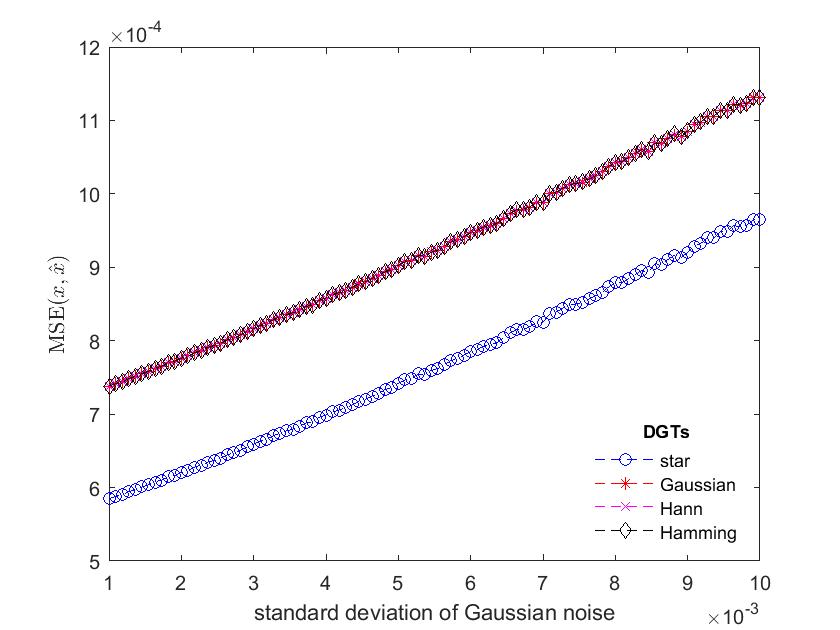}
         \caption{5338-284437-0025, $(L,a,b)=(29835,17,13)$, Gaussian noise}
         \label{5338}
     \end{subfigure}\hfill
     \begin{subfigure}[b]{0.4\textwidth}
         \centering
         \includegraphics[width=\textwidth]{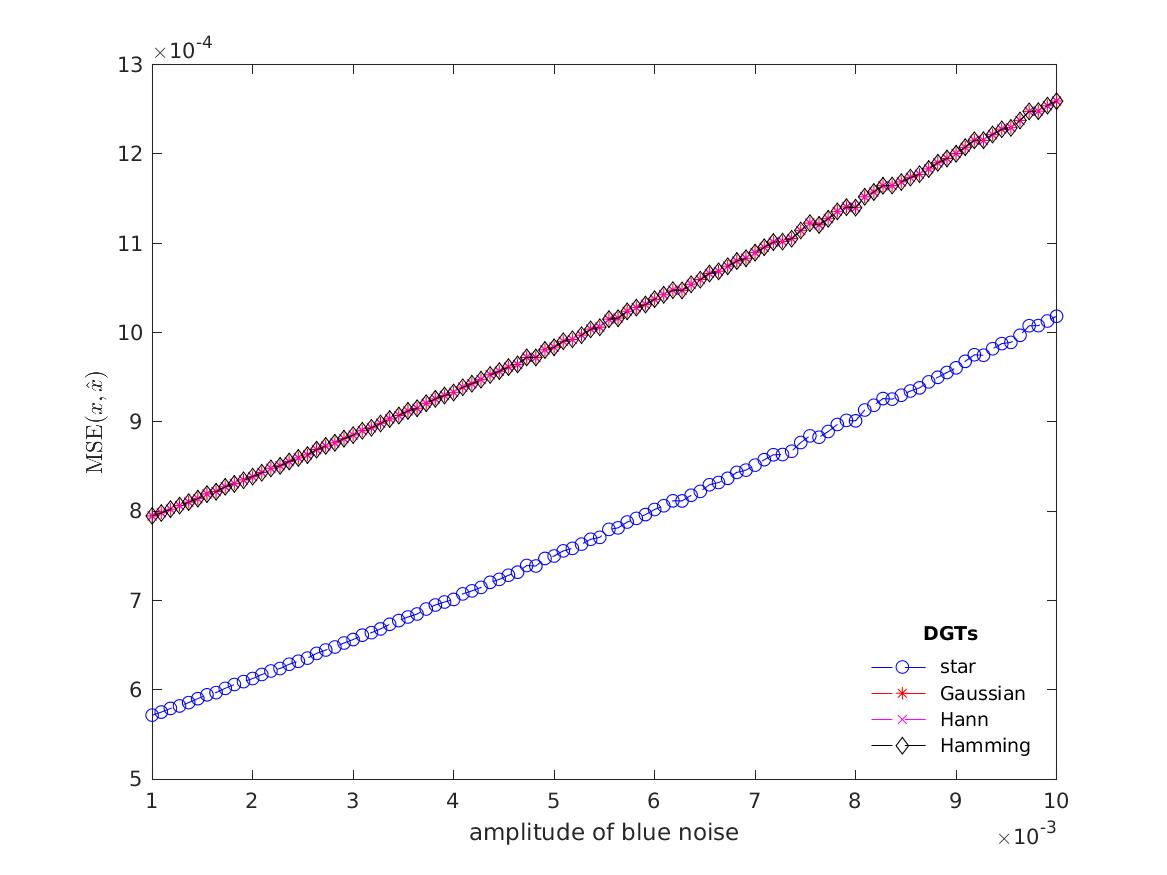}
         \caption{5338-284437-0025, $(L,a,b)=(29835,17,13)$, blue noise}
         \label{5338blue}
     \end{subfigure}
     \hfill
     \begin{subfigure}[b]{0.4\textwidth}
         \centering
         \includegraphics[width=\textwidth]{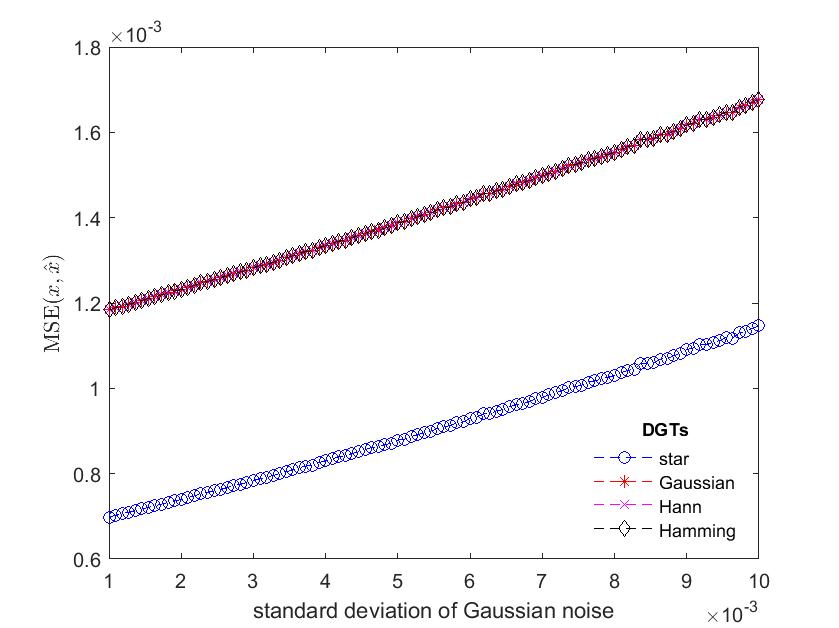}
         \caption{3752-4944-0042, $(L,a,b)=(51051,33, 17)$, Gaussian noise}
         \label{375242}
     \end{subfigure}
     \hfill
     \begin{subfigure}[b]{0.4\textwidth}
         \centering
         \includegraphics[width=\textwidth]{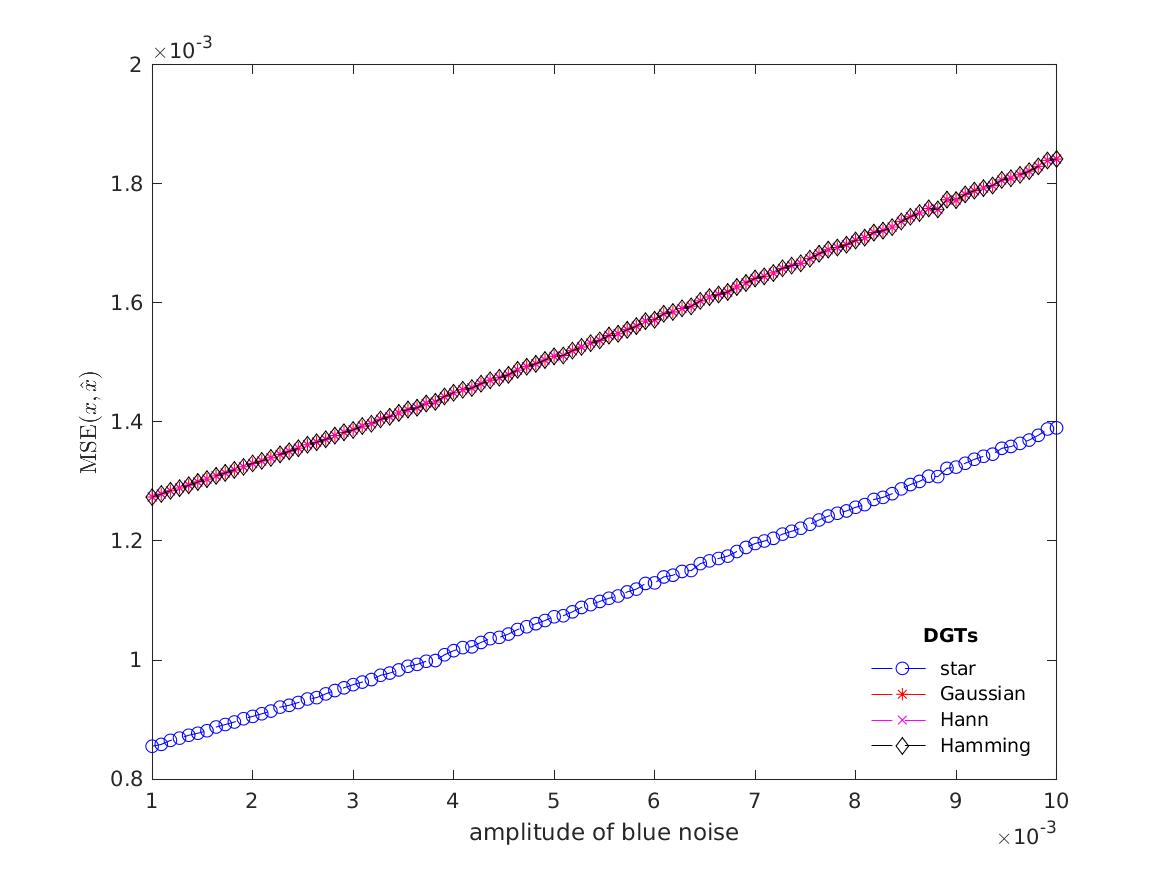}
         \caption{3752-4944-0042, $(L,a,b)=(51051,33, 17)$, blue noise}
         \label{375242blue}
     \end{subfigure}\hfill
     \begin{subfigure}[b]{0.4\textwidth}
         \centering
         \includegraphics[width=\textwidth]{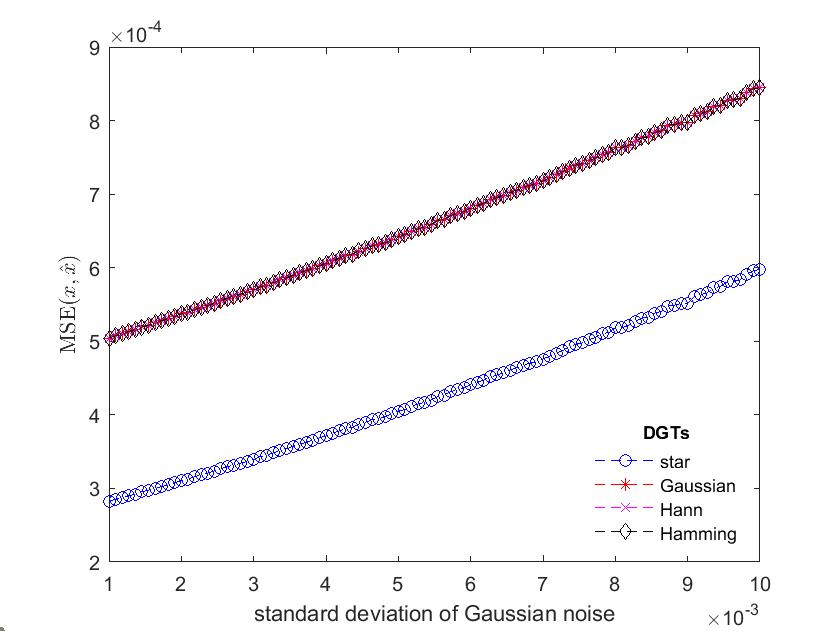}
         \caption{5694-64038-0013, $(L,a,b)=(51051,33, 17)$, Gaussian noise}
         \label{5694}
     \end{subfigure}\hfill
     \begin{subfigure}[b]{0.4\textwidth}
         \centering
         \includegraphics[width=\textwidth]{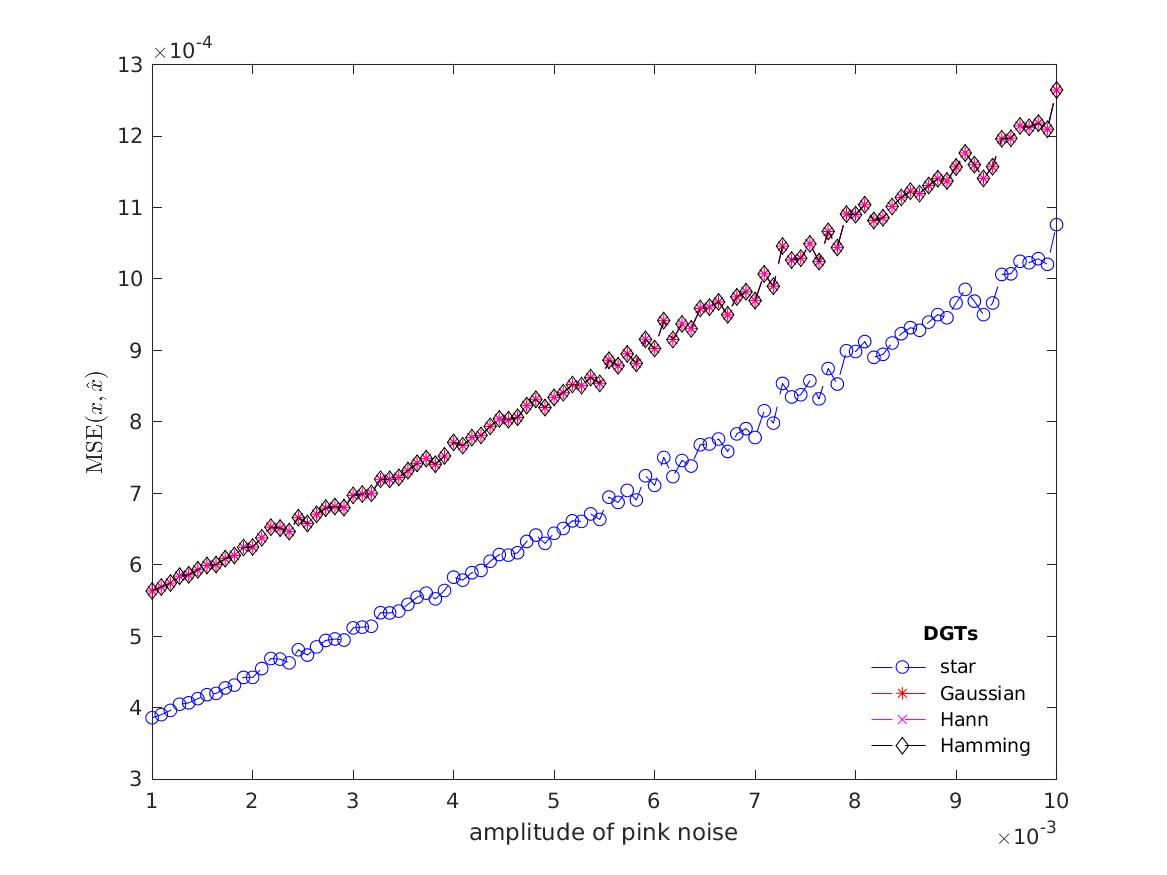}
         \caption{5694-64038-0013, $(L,a,b)=(51051,33, 17)$, pink noise}
         \label{5694pink}
     \end{subfigure}
     \captionsetup{justification=centering}
        \caption{Rate of denoising success for 4 speech signals with different parameters $(L,a,b)$, contaminated by Gaussian (left) and coloured (right) noise. Note that 3 of the 4 methods roughly coincide.}
        \label{gaussian2}
\end{figure}

\begin{figure}
     \centering
     \begin{subfigure}[b]{0.4\textwidth}
         \centering         \includegraphics[width=\textwidth]{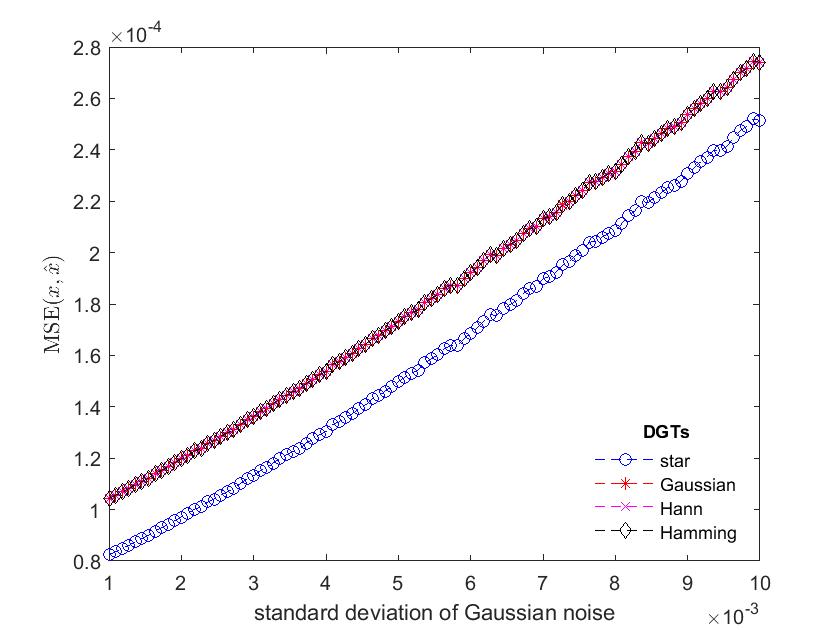}
         \caption{5895-34615-0001, $(L,a,b)=(51051,33,17)$, Gaussian noise}
         \label{5895}
     \end{subfigure}
     \hfill
     \begin{subfigure}[b]{0.4\textwidth}
         \centering         \includegraphics[width=\textwidth]{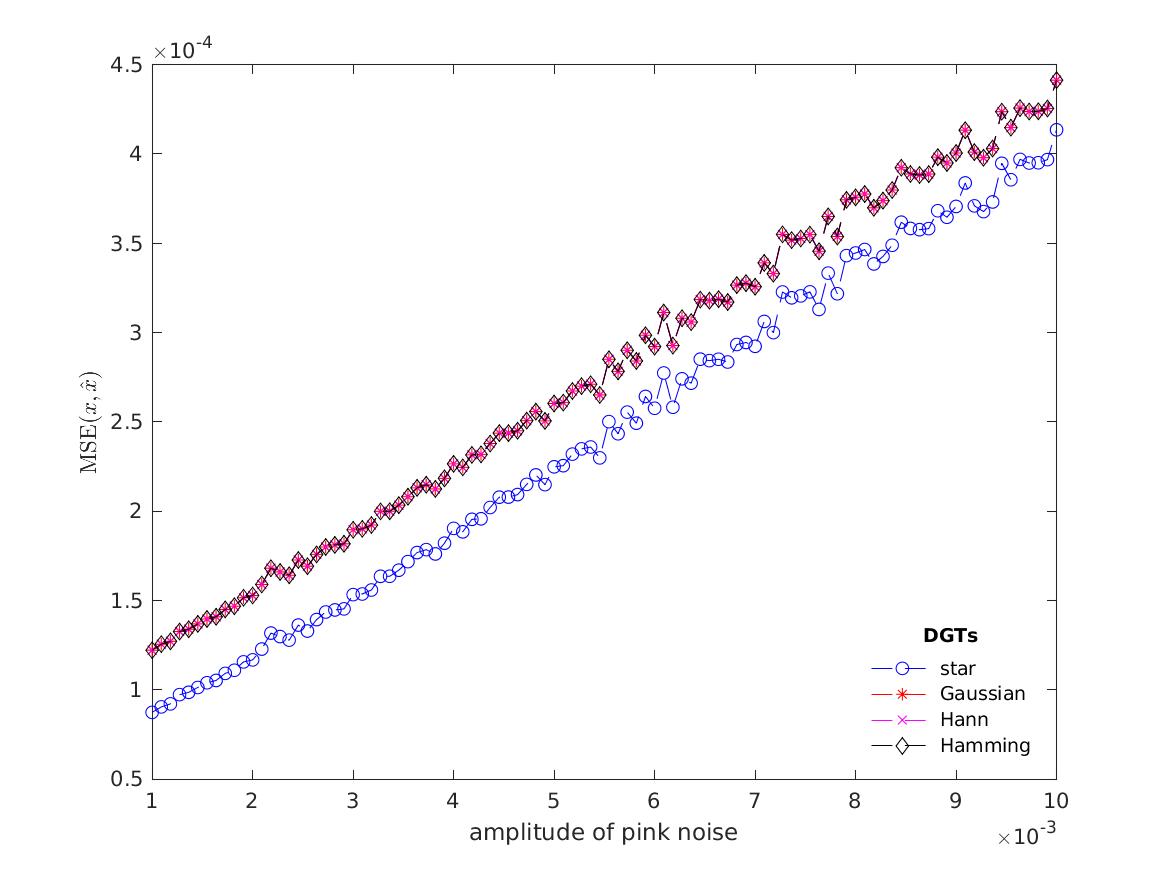}
         \caption{5895-34615-0001, $(L,a,b)=(51051,33,17)$, pink noise}
         \label{5895pink}
     \end{subfigure}
     \hfill
     \begin{subfigure}[b]{0.4\textwidth}
         \centering
         \includegraphics[width=\textwidth]{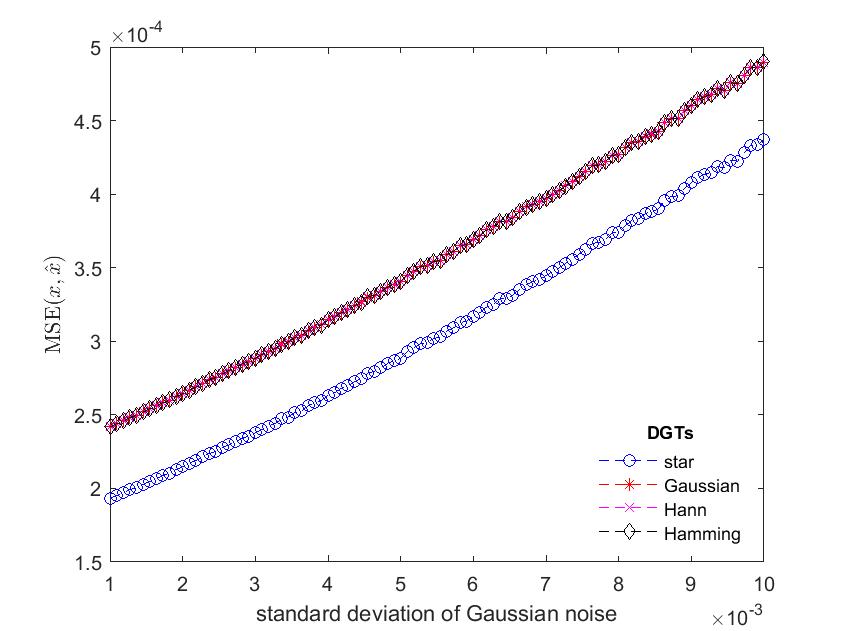}
         \caption{2428-83699-0035, $(L,a,b)=(41769,21,17)$, Gaussian noise}
         \label{2428}
     \end{subfigure}\hfill
     \begin{subfigure}[b]{0.4\textwidth}
         \centering
         \includegraphics[width=\textwidth]{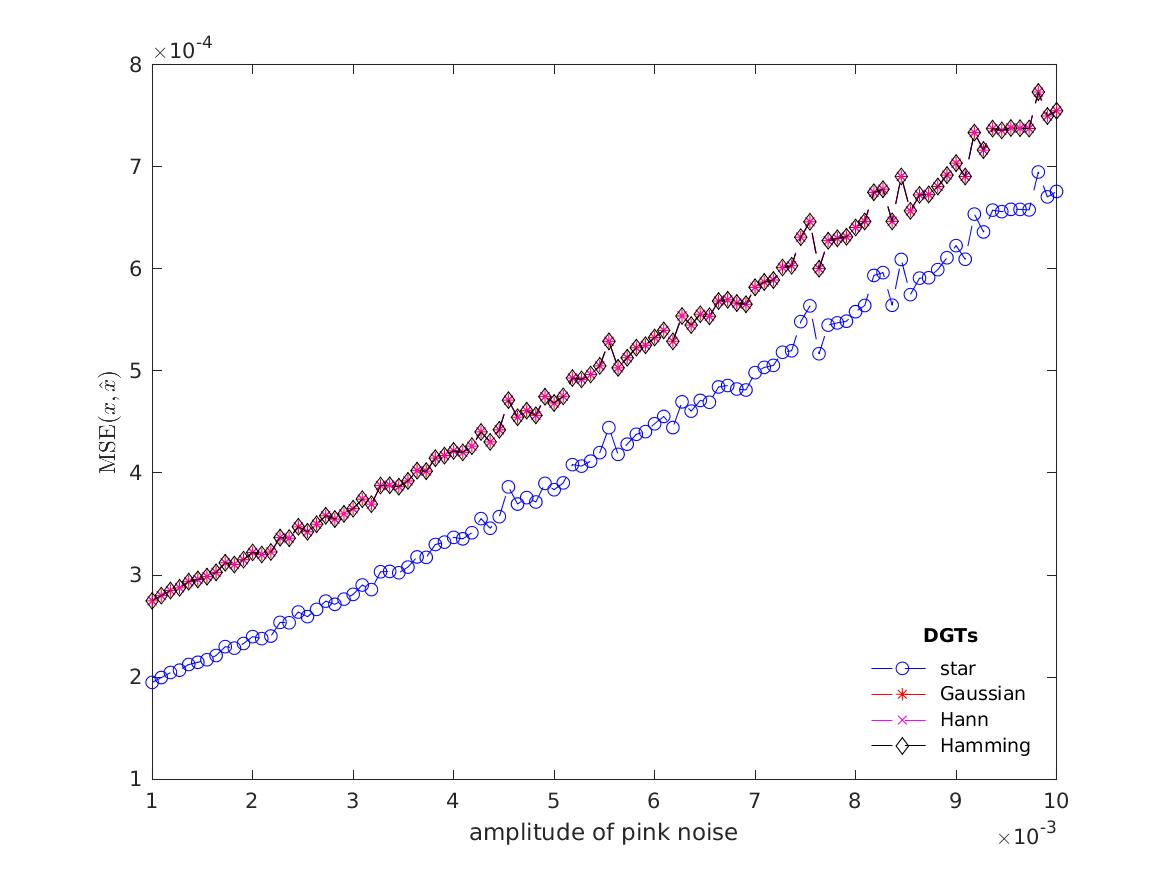}
         \caption{2428-83699-0035, $(L,a,b)=(41769,21,17)$, pink noise}
         \label{2428pink}
     \end{subfigure}\hfill
     \hfill
     \begin{subfigure}[b]{0.4\textwidth}
         \centering
         \includegraphics[width=\textwidth]{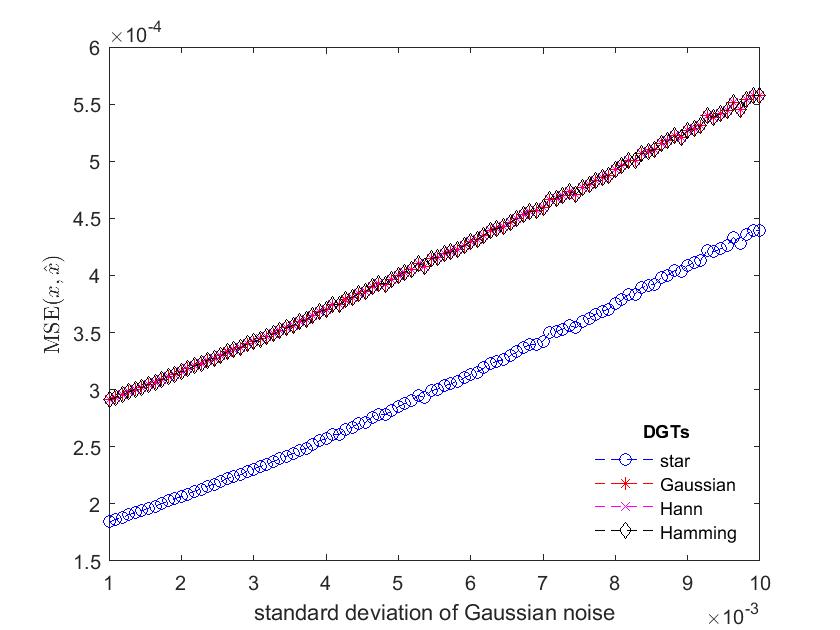}
         \caption{2803-154320-0006, $(L,a,b)=(33915,51, 19)$, Gaussian noise}
         \label{2803}
     \end{subfigure}
     \hfill
     \begin{subfigure}[b]{0.4\textwidth}
         \centering
         \includegraphics[width=\textwidth]{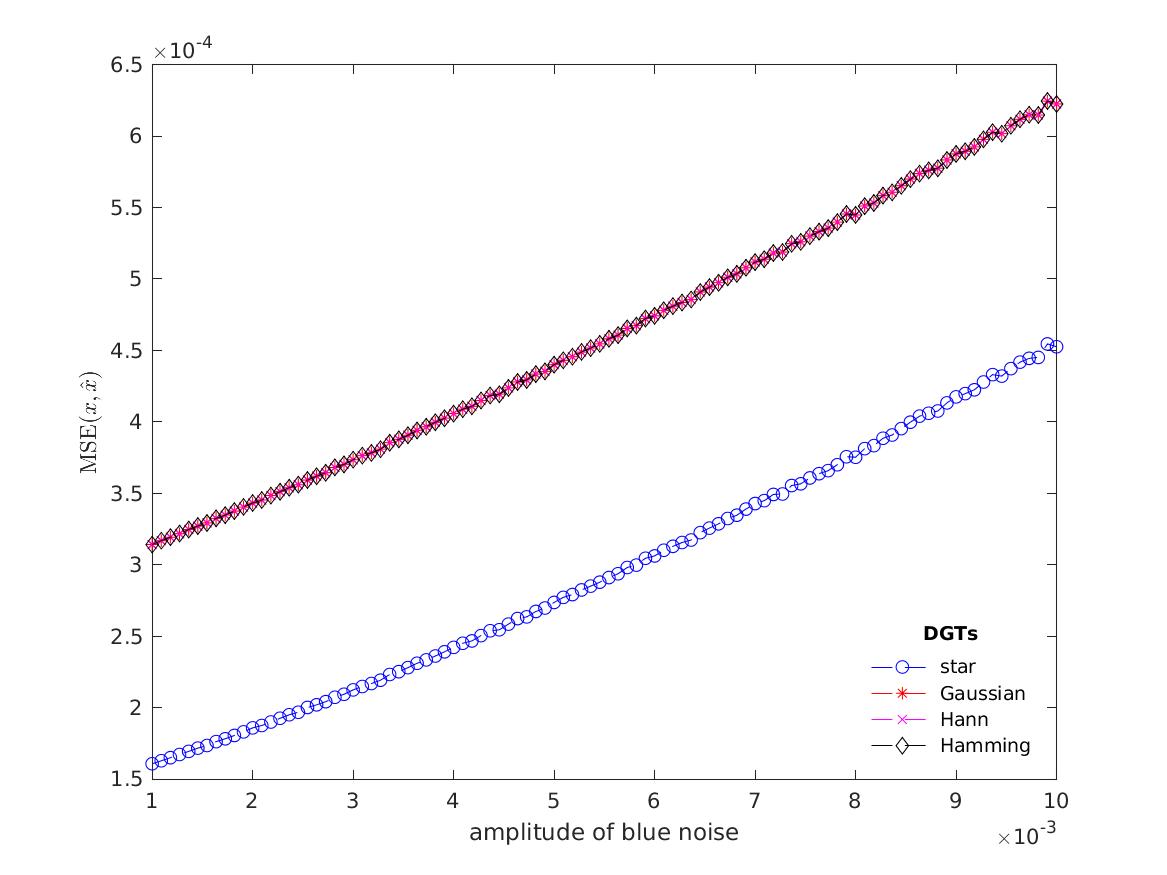}
         \caption{2803-154320-0006, $(L,a,b)=(33915,51, 19)$, blue noise}
         \label{2803blue}
     \end{subfigure}
     \hfill
     \begin{subfigure}[b]{0.4\textwidth}
         \centering
         \includegraphics[width=\textwidth]{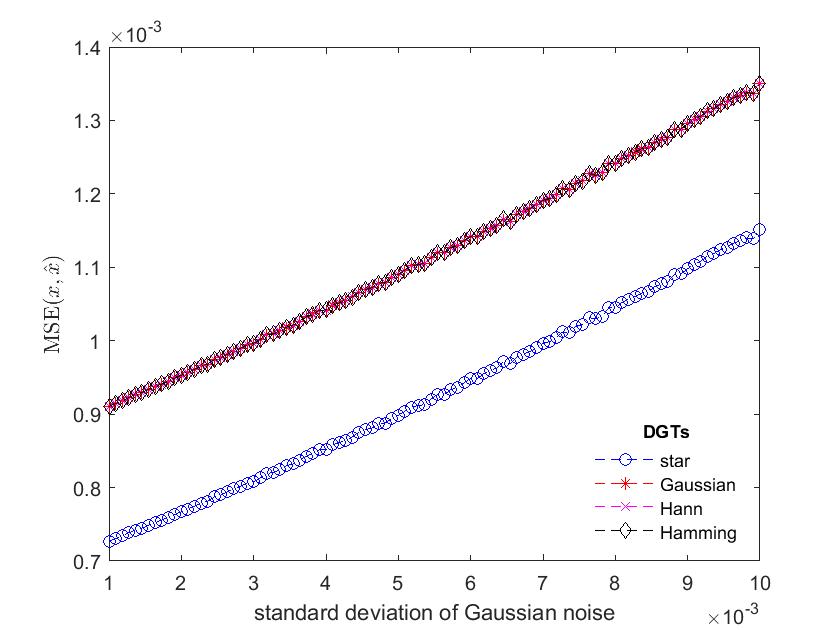}
         \caption{3752-4944-0008, $(L,a,b)=(29835,17, 13)$, Gaussian noise}
         \label{37528}
     \end{subfigure}\hfill
     \begin{subfigure}[b]{0.4\textwidth}
         \centering
         \includegraphics[width=\textwidth]{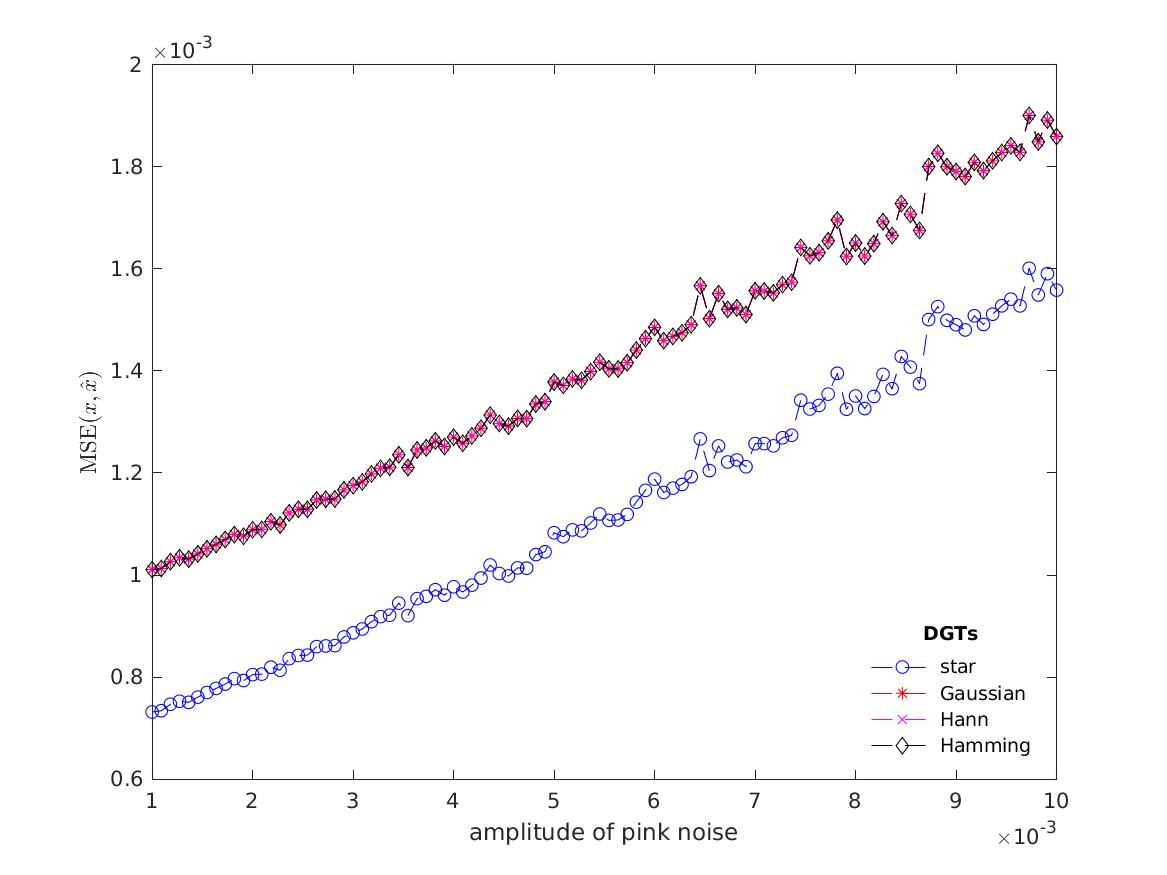}
         \caption{3752-4944-0008, $(L,a,b)=(29835,17, 13)$, pink noise}
         \label{37528pink}
     \end{subfigure}
     \captionsetup{justification=centering}
        \caption{Rate of denoising success for 4 speech signals with different parameters $(L,a,b)$, contaminated by Gaussian (left) and coloured (right) noise. Note that 3 of the 4 methods roughly coincide.}
        \label{gaussian3}
\end{figure}

\section{Experiments and Results}
We present two sets of experiments in the following subsections.
\subsection{Fixed $(L,a,b)$ with varying $\sigma$}
We fix for each of the 12 signals the lattice parameters $a,b$, with respect to each artificial dimension $L$. We add to all signals zero-mean Gaussian noise with varying standard deviation, using the vector $\bm{\sigma}$, and perform analysis denoising for each entry of $\bm{\sigma}$. For the coloured noise cases, we randomly split the set of 12 signals into two subsets, of 6 signals each. We add to the signals of the first and second subset blue and pink noise respectively, with varying amplitude controlled by the vector $\bm{\sigma}$, and perform analysis denoising for each entry of $\bm{\sigma}$. The left column in Fig.~\ref{gaussian1}-\ref{gaussian3} demonstrates for different signals, how the 4 resulting MSEs scale in the case of the Gaussian noise, as its standard deviation increases. Clearly, our proposed DGT outperforms the rest of DGTs, consistently for all signals and for different choices of artificial dimension with time-frequency parameters. Similarly, the right column in Fig.~\ref{gaussian1}-\ref{gaussian3} demonstrates for different signals, how the 4 resulting MSEs scale in the case of blue and pink noise, as the scaling factor of each coloured noise's amplitude increases. We observe that star-DGT is more robust than the rest of DGTs, even when the amplitude of each coloured noise is almost equal to the amplitude of the speech signal to which it is added.
\subsection{Fixed $L$ and $\sigma$, with varying $(a,b)$}
We randomly pick 6 out of the 12 speech signals (we prefer to examine signals with different artificial dimensions). We alter for each signal the time-frequency parameters $a,b$ with respect to its artificial dimension. We consider the fixed scalar $\sigma=0.001$ serving as both the standard deviation of the Gaussian noise and the scaling factor controlling the coloured noises' amplitude. For different pairs of $(a,b)$, we add Gaussian noise to all six signals, blue noise to three of the six signals and pink noise to the rest of them. Finally, we denoise all signals for all types of noise and present the resulting MSEs in Tables~\ref{251par}-\ref{8842par}. For all choices of $(a,b)$, star-DGT (indicated in bold in each subtable) outperforms the baseline DGTs, consistently for all signals, for both Gaussian and coloured noises. Additionally, we see that among all examined pairs of lattice parameters, star-DGT achieves the smallest MSE (indicated in purple in each subtable) when $a,b$ are chosen as the two largest factors in the prime factorization of $L$; the rest of DGTs do not seem to benefit much from this selection. On the other hand, among all examined choices of $(a,b)$, star-DGT performs slightly worse when $a=b$. For example, as indicated in Tables~\ref{5694par} and \ref{8842par}, star-DGT reaches a slightly bigger MSE when $a=b=11$ and $a=b=3^2$, respectively.

\section{Conclusion and Future Directions}
In the present paper, we took advantage of a window vector to generate a spark deficient Gabor frame and introduced a redundant analysis Gabor operator/DGT, namely the star-DGT, associated with this SDGF. We then applied the star-DGT to analysis denoising, along with three other DGTs generated by state-of-the-art window vectors in the field of Gabor Analysis. First, we fixed the ambient dimension and the time-frequency parameters, and altered the standard deviation of the Gaussian noise and the amplitude of the coloured noises. Second, we examined how different pairs of lattice parameters, with fixed standard deviation and amplitude of the Gaussian and coloured noises respectively, affect the performance of analysis denoising. All experiments confirm improved robustness: the increased amount of linear dependencies provided by this SDGF, yields for all speech signals a lower MSE for the proposed method. Future directions will include the combination of the present framework with deep learning methods \cite{gabdl}, as well as the examination of the robustness of the weighted combination
\begin{equation}
    \begin{split}
        \text{minimize}\quad&\|\Phi x\|_1+\|x\|_{\mathrm{TV}}+\frac{\mu}{2}\|x-x_0\|_2^2\\
        \text{subject to}\quad &\|x-y\|_2\leq \eta,
    \end{split}
\end{equation}
where $\|x\|_{\mathrm{TV}}$ is the total-variation norm of $x$.
Finally, it would be interesting to develop a rigorous mathematical proof, explaining why star-DGT benefits more when the lattice parameters are chosen as the two largest primes in the prime factorization of a signal's dimension.

\section*{Acknowledgments}
V. Kouni would like to thank G. Paraskevopoulos for his valuable advice and insightful discussions around the framework presented in this paper.

\bibliography{sn-bibliography}


\end{document}